\documentclass[%
 aip,
 amsmath,amssymb,
 reprint,%
]{revtex4-1}

\usepackage{textcomp}
\usepackage{soul}       
\setstcolor{red}        

\usepackage{lipsum}
\usepackage{graphicx}
\usepackage{dcolumn}
\usepackage{bm}
\usepackage{xcolor}
\usepackage[utf8]{inputenc}
\usepackage[T1]{fontenc}
\usepackage{mathptmx}
\usepackage{etoolbox}
\usepackage{placeins}
\usepackage{siunitx}
\makeatletter
\def\@email#1#2{%
 \endgroup
 \patchcmd{\titleblock@produce}
  {\frontmatter@RRAPformat}
  {\frontmatter@RRAPformat{\produce@RRAP{*#1\href{mailto:#2}{#2}}}\frontmatter@RRAPformat}
  {}{}
}%
\makeatother

\makeatletter
\def\@email#1#2{%
 \endgroup
 \patchcmd{\titleblock@produce}
  {\frontmatter@RRAPformat}
  {\frontmatter@RRAPformat{\produce@RRAP{*#1\href{mailto:#2}{#2}}}\frontmatter@RRAPformat}
  {}{}
}%
\makeatother

\makeatletter
\renewcommand\section{\@startsection {section}{1}{\z@}%
  {-1.5ex \@plus -1ex \@minus -.2ex}%
  {0.5ex \@plus.2ex}%
  {\normalfont\normalsize\bfseries\raggedright}}
\renewcommand\subsection{\@startsection {subsection}{2}{\z@}%
  {-1.25ex\@plus -1ex \@minus -.2ex}%
  {0.3ex \@plus .2ex}%
  {\normalfont\normalsize\bfseries\raggedright}}
\makeatother

\begin{document}
\preprint{AIP/123-QED}


\author{Giulia Meucci\textsuperscript{*,}}
\affiliation{Department of Energy Conversion and Storage, Technical University of Denmark, 2800 Kgs.Lyngby, Denmark}  \email{giuliam@dtu.dk}

\author{Thies Jansen}
\affiliation{Department of Energy Conversion and Storage, Technical University of Denmark, 2800 Kgs.Lyngby, Denmark}

\author{Pinelopi O. Konstantinopoulou}
\affiliation{Department of Energy Conversion and Storage, Technical University of Denmark, 2800 Kgs.Lyngby, Denmark}

\author{Dags Ol{\v{s}}teins}
\affiliation{Department of Energy Conversion and Storage, Technical University of Denmark, 2800 Kgs.Lyngby, Denmark}

\author{Damon J. Carrad}
\affiliation{Department of Energy Conversion and Storage, Technical University of Denmark, 2800 Kgs.Lyngby, Denmark}

\author{Carina Babu Maliakkal}
\affiliation{Catalan Institute of Nanoscience and Nanotechnology - ICN2 (CSIC and BIST), Campus UAB, Bellaterra, Barcelona, 08193, Catalonia, Spain}

\author{Jordi Arbiol}
\affiliation{Catalan Institute of Nanoscience and Nanotechnology - ICN2 (CSIC and BIST), Campus UAB, Bellaterra, Barcelona, 08193, Catalonia, Spain}\affiliation{ICREA, Pg. Lluís Companys 23, 08010 Barcelona, Catalonia, Spain}

\author{Nicolas Bergeal}
\affiliation{Laboratoire de Physique et d’Etude des Matériaux, ESPCI Paris, Université PSL, CNRS, Sorbonne Université, Paris, France}

\author{Manuel Bibes}
\affiliation{Laboratoire Albert Fert, CNRS, Thales, Université Paris-Saclay, 91767 Palaiseau, France}

\author{Roberta Citro}
\affiliation{Dipartimento di Fisica “E.R. Caianiello,” Università di Salerno, 84084 Fisciano, Italy} \affiliation{CNR-SPIN, c/o Universita di Salerno, 84084 Fisciano, Italy}

\author{Emiliano Di Gennaro}
\affiliation{Dipartimento di Fisica "Ettore Pancini", Università degli Studi di Napoli “Federico II”,  Complesso di Monte S. Angelo, Via Cinthia, 80126 Naples, Italy}

\author{Alexei Kalaboukhov}
\affiliation{Department of Microtechnology and Nanoscience (MC2), Chalmers University of Technology, 41296 Gothenburg, Sweden}

\author{Nicola Manca}
\affiliation{CNR-SPIN, C.so F. M. Perrone 24, 16152 Genova, Italy}

\author{Carmine A. Perroni}
\affiliation{CNR-SPIN,  Complesso di Monte S. Angelo, via Cinthia, 80126  Naples, Italy}
\affiliation{Dipartimento di Fisica "Ettore Pancini", Università degli Studi di Napoli “Federico II”,  Complesso di Monte S. Angelo, Via Cinthia, 80126 Naples, Italy}

\author{Marco Salluzzo}
\affiliation{CNR-SPIN,  Complesso di Monte S. Angelo, via Cinthia, 80126  Naples, Italy}

\author{Pawe{\l} W{\'o}jcik}
\affiliation{AGH University of Krakow, Faculty of Physics and Applied Computer Science, al. A. Mickiewicza 30, 30-059 Krakow, Poland}

\author{Felix Trier}
\affiliation{Department of Energy Conversion and Storage, Technical University of Denmark, 2800 Kgs.Lyngby, Denmark}

\author{Nini Pryds}
\affiliation{Department of Energy Conversion and Storage, Technical University of Denmark, 2800 Kgs.Lyngby, Denmark}

\author{Fabio Miletto Granozio}
\affiliation{CNR-SPIN, Complesso di Monte S. Angelo, via Cinthia, 80126  Naples, Italy}

\author{Alessia Sambri}
\affiliation{CNR-SPIN,  Complesso di Monte S. Angelo, via Cinthia, 80126  Naples, Italy}

\author{Thomas S. Jespersen}
\affiliation{Department of Energy Conversion and Storage, Technical University of Denmark, 2800 Kgs.Lyngby, Denmark}\affiliation{Center For Quantum Devices, Niels Bohr Institute, University of Copenhagen, 2100 Copenhagen, Denmark}

\date{\today}

\title{A Negative-$U$ Double Quantum Dot in a LaAlO$_3$/SrTiO$_3$ Membrane on Silicon}

\begin{abstract}
The development of freestanding oxide membranes creates new opportunities to integrate oxide functionalities with silicon technology. A central question is whether local gates can control the correlated electronic states of transferred membranes at the scale required for hybrid mesoscopic quantum devices. Here, we demonstrate local electrostatic control of freestanding LaAlO$_3$/SrTiO$_3$ (LAO/STO) membranes transferred onto silicon. In a dual-gated constriction in the interfacial electron system, we observe an oxide-based double quantum dot (DQD) in the regime of attractive on-site interactions ($U<0$).  Low-temperature transport reveals the characteristic honeycomb charge stability diagram, and spectroscopy identifies a paired-electron ground state at low magnetic fields, with odd-occupancy states becoming accessible above the pairing field. The observed transport spectra are reproduced by simulations within an attractive two-site Hubbard model, and we extract the relevant DQD parameters, including intradot and interdot interactions, interdot tunnel coupling, and spin-orbit-assisted tunnel coupling. These findings establish oxide membranes as hosts for DQDs and extend coupled quantum dot systems into the regime of attractive on-site interactions.
\end{abstract}

\maketitle
\begin{figure*}[!t]
\centering
\includegraphics[width = \linewidth]{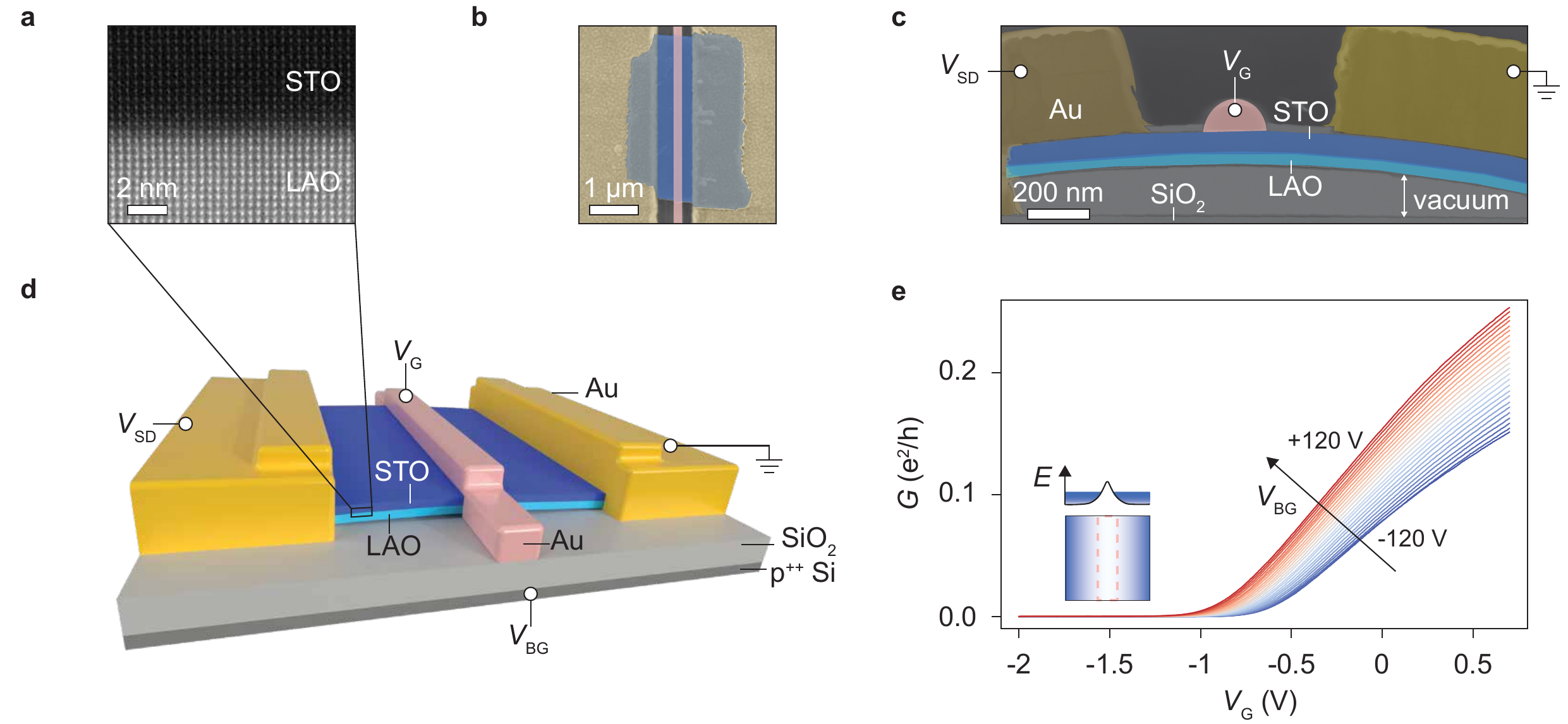}
\caption{ \textbf{Local electrostatic control of freestanding oxide membranes.}
\textbf{a} Cross-sectional atomic resolution HAADF STEM image of the LAO/STO interface in a freestanding membrane.
\textbf{b} Top-view and \textbf{c} cross-sectional false-color scanning electron micrographs of two transferred LAO/STO micro-membrane devices with Ti/Au contacts and a local top-gate on a Si/SiO$_2$ substrate.
\textbf{d} Schematic of the device architecture showing the single local top-gate and global back-gate.
\textbf{e} Room-temperature conductance, $G$, as a function of a single local-gate voltage, $V_\text{G}$, for back-gate voltages, $V_\text{BG}$, ranging from $-120$ to $+120$ V, measured in a device with top-gate width 100 nm. The inset highlights the low-$V_\text{G}$ regime, where the top-gate locally depletes the carrier density, leading to pinch-off. The blue shading indicates the local carrier density in the interfacial 2DEG, with lighter shades corresponding to lower carrier densities.
}
\end{figure*}

\begin{figure*}[htb]
\centering
\includegraphics[width = \linewidth]{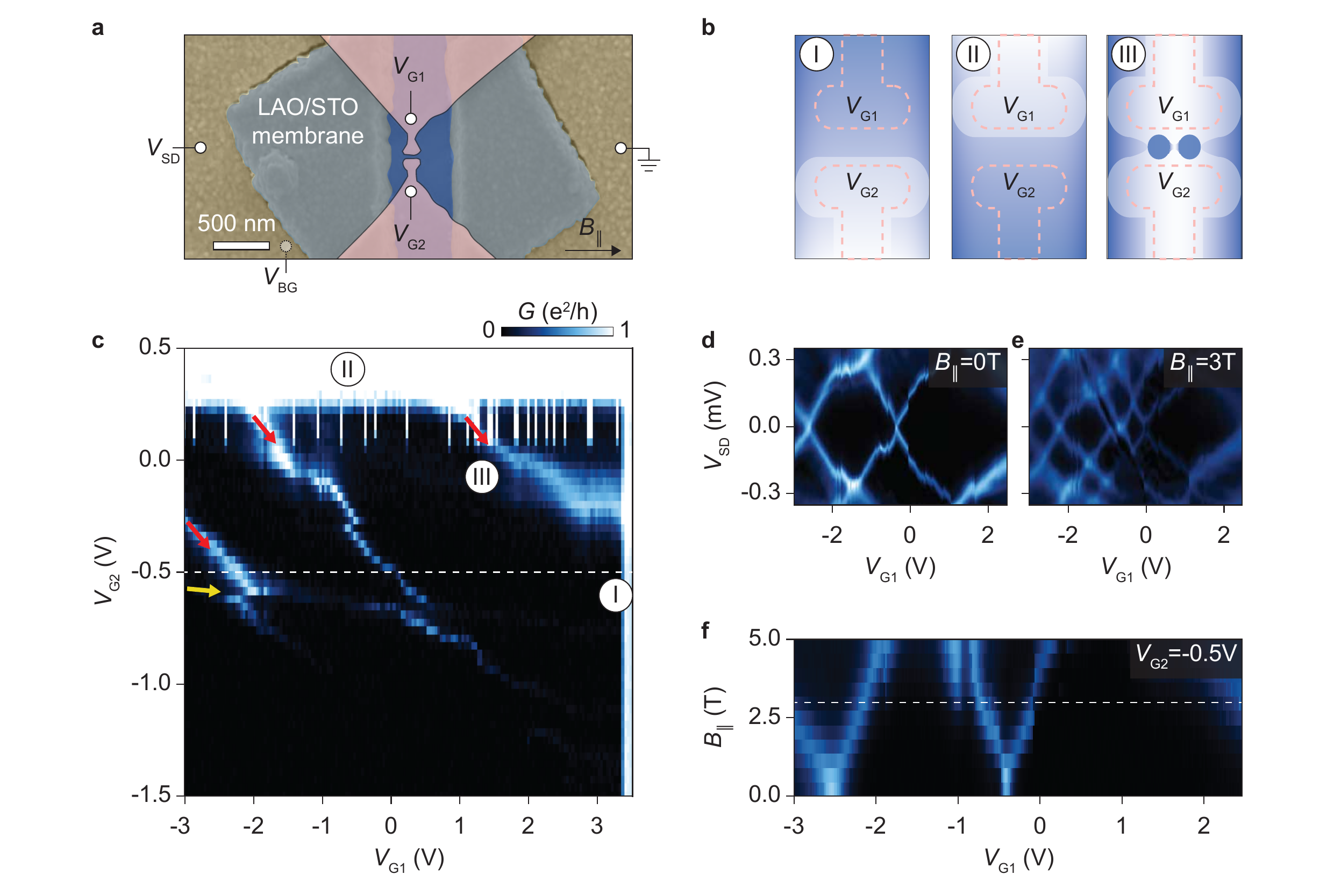}
\caption{ \textbf{Negative-$U$ QDs in a gated constriction.}
\textbf{a} False-color scanning electron micrograph of the device and schematic of the measurement setup. The back-gate voltage is applied through the back side of the substrate, as shown in Fig.~1d.
\textbf{b} Schematic illustration of the three transport regimes realized by tuning  the local-gates. Blue indicates the local carrier density in the interfacial 2DEG, with lighter shades corresponding to lower carrier density. The positions of the QDs in regime III are schematic and not experimentally determined.
\textbf{c} Differential conductance, $G$, as a function of local-gate voltages, $V_{\text{G}1}$ and $V_{\text{G}2}$ at zero source-drain bias, $V_{\mathrm{SD}}=0$.   The white dashed line indicates $V_{\mathrm{G}2}=-0.5$ V and corresponds to the line cuts shown in \textbf{d-f}. Red and yellow arrows indicate the two families of Coulomb peaks. Roman numerals correspond to the regimes identified in \textbf{b}.
 \textbf{d}   $G$, as a function of $V_{\text{G}1}$ and $V_{\mathrm{SD}}$, at fixed $V_{\text{G}2}=-0.5$ V and $B_{\parallel}=0$.  
\textbf{e}  $G$, as a function of $V_{\text{G}1}$ and $V_{\mathrm{SD}}$, at fixed $V_{\text{G}2}=-0.5$ V and $B_{\parallel}=3$ T.
\textbf{f}  $G$, as a function of $V_{\mathrm{G}1}$ and the in-plane magnetic field $B_{\parallel}$ at fixed $V_{\mathrm{G}2}=-0.5$ V, $V_{\mathrm{SD}}=0$. The white dashed line marks $B_{\parallel}=3$ T, corresponding to the line cut shown in \textbf{e}. 
Measurements were performed at $V_\mathrm{BG}=-7$ V, $T_\mathrm{base}=20$ mK, and $B_\perp=200$ mT.
}
\end{figure*}

\begin{figure*}[hbt]
\centering
\includegraphics[width = \linewidth]{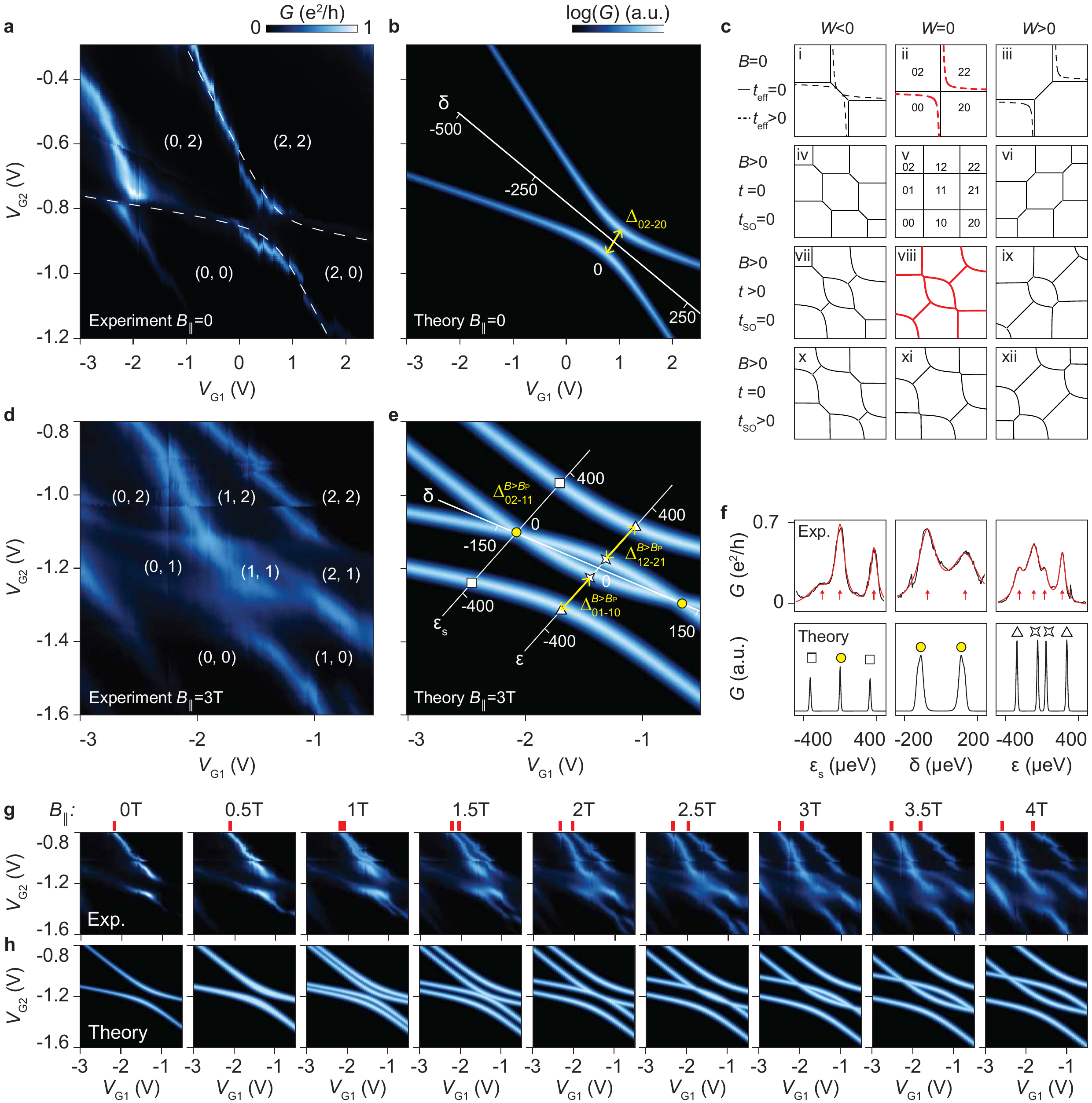}
\caption{ \textbf{Negative-$U$ DQD.}
\textbf{a} Experimental differential conductance, $G$, versus $V_{\mathrm{G}1}$ and $V_{\mathrm{G}2}$ at $B_\parallel=0$ T. Numbers denote charge states $(N_1,N_2)$ relative to unknown even offsets. White dashed lines show fits to the charge-transition branches.
\textbf{b} Simulated $G$ at $B_\parallel=0$ T. The white line marks the detuning axis, $\delta$ (markers in \textmu eV). The yellow line shows $\Delta_{02-20}$.
\textbf{c} Negative-$U$ DQD stability diagrams for different combinations of $W$, $t$, $t_{\mathrm{SO}}$, and $B$.
\textbf{d} Experimental $G$ at $B_\parallel=3$ T. 
\textbf{e} Simulated $G$ at $B_\parallel=3$ T. White lines indicate the detuning, $\delta$, energy, $\epsilon$, and shifted-energy, $\epsilon_\mathrm{S}$, axes (markers in \textmu eV). The yellow lines and marker indicate $\Delta_{01 - 10}^{B>B_\mathrm{P}}$, $\Delta_{12-21}^{B>B_\mathrm{P}}$  and $\Delta_{02-11}^{B>B_\mathrm{P}}$.
\textbf{f} Line profiles along $\epsilon_\mathrm{S}$, $\delta$, and $\epsilon$ for the experiment (top; red: Lorentzian fits) and simulation (bottom).
\textbf{g,h} Experimental (\textbf{g}) and simulated (\textbf{h}) evolution of $G$ with $B_\parallel$. }
\end{figure*}

\section*{INTRODUCTION}
\indent The development of synthesis and transfer techniques for freestanding complex oxide membranes represents a potential breakthrough for integrating oxide functionalities into semiconductor electronics and quantum-device architectures\cite{lu2016synthesis,lu2019freestanding,kum2019epitaxial,sambri2020self,eom2021electronically, choo2024oxide}. 
Complex oxides exhibit a rich interplay among charge, spin, orbital, and lattice degrees of freedom, resulting in electronic, magnetic, and superconducting phases absent in conventional semiconductors and of both fundamental and technological interest\cite{hwang2012emergent,ramesh2019creating}. However, these functionalities are highly sensitive to crystal defects, doping, strain, and other perturbations, and reproducible performance has traditionally required optimized epitaxial growth on specific lattice-matched substrates that are generally incompatible with semiconductor technologies and processing. Similar to the transfer of exfoliated 2D materials\cite{novoselov2004electric} and 1D semiconductor nanowires\cite{jia2019nanowire}, freestanding oxide membranes may overcome these limitations by decoupling material synthesis from device fabrication and integration\cite{lu2016synthesis,lu2019freestanding,kum2019epitaxial,sambri2020self, eom2021electronically}.\\ 
\indent While structural transfer methods are well established, it remains unclear whether membrane formation preserves the correlated-electron functionalities of oxide heterostructures and whether transferred membranes retain the local electrostatic tunability essential for electronic devices and the engineering of quantum confinement. Here, we address these questions by studying low-temperature transport in locally gated devices fabricated from spalled LAO/STO micro-membranes\cite{sambri2020self} transferred onto silicon substrates with the STO layer facing up.  The interfacial 2DEG in LAO/STO is of particular interest as numerous previous studies of epitaxial heterostructures have shown a combination of rich correlated-electron phenomena\cite{ohtomo2004high,  reyren2007superconducting, christensen2019strain}, atypical manifestation of the quantum Hall effect\cite{matsubara2016observation, trier2016quantization}, gate tunability\cite{thiel2006tunable, caviglia2008electric, caviglia2010tunable},  and unconventional electron pairing described by a effective attractive on-site interaction, $U<0$\cite{cheng2015electron,tomczyk2016micrometer,prawiroatmodjo2017transport,annadi2018quantized,bjorlig2020g,briggeman2020pascal}.  \\
\indent   We demonstrate local electrostatic control of transferred membranes by fabricating and studying membrane field-effect transistors with local top-gates, acting through the STO layer, as narrow as 100 nm. Furthermore, in the tunneling regime near pinch-off of a double-gated device, we observe a double quantum dot (DQD), a previously unexplored system in oxide materials. The device preserves the attractive on-site interactions previously observed in single quantum dots (QDs) formed in conventional LAO/STO heterostructures\cite{cheng2015electron,tomczyk2016micrometer,prawiroatmodjo2017transport,annadi2018quantized,bjorlig2020g,briggeman2020pascal,meucci2026intrinsic}, thereby realizing a unique negative-$U$ counterpart to conventional DQDs. This attractive interaction reverses the usual two-electron energetics of a DQD: instead of favoring spatially separated single-electron occupation, the ground state consists of local pairs. Transport spectroscopy and modeling show that the device realizes an attractive regime in which the ground state at zero detuning is dominated by a hybridized configuration of localized electron pairs. Unlike paired states induced in conventional systems through proximity to superconductors\cite{de2010hybrid}, the negative-$U$ systems retain a discrete paired spectrum without relying on a superconducting condensate or a quasiparticle continuum above the gap.

\section*{RESULTS}
\subsection*{Local Electrostatic Control}
Following Ref.\ \citenum{sambri2020self}, freestanding LAO/STO micro-membranes (LAO/STO thickness: 55~nm~/~55~nm) were realized by strain-induced spalling. This method provides precise control over membrane position and lateral dimensions (2 to 20~$\mu$m)\cite{dahm2021size}, enables transfer onto arbitrary substrates, and preserves high-quality epitaxial interfaces\cite{sambri2020self}, as demonstrated by the atomic resolution aberration corrected high angle annular dark field (HAADF) scanning transmission electron microscopy (STEM) image in Fig.~1a. \\
\indent Micro-membranes were transferred, with the STO layer facing upward, onto p$^{++}$-doped Si substrates, capped with a 500-nm-thick SiO$_2$ insulating layer. Ohmic Ti/Au contacts to the interfacial 2DEG were fabricated in a two-terminal configuration by shallow-angle milling followed by metal deposition\cite{erlandsen2022two} (see Materials and Methods). To achieve local electrostatic control, narrow top-gates were defined by electron-beam lithography. The doped Si substrate serves as a global back-gate, enabling direct comparison of local and global electrostatic control. Representative scanning electron micrographs of devices with a single local top-gate of width 200~nm are shown in Fig.~1b,c, and a schematic of the device architecture is presented in Fig.~1d.\\
\indent Figure 1e shows the room-temperature conductance, $G$, of a representative device with top-gate width 100 nm, as a function of the local-gate potential, $V_\mathrm{G}$, for several values of the back-gate potential, $V_\mathrm{BG}$. Additional measurements from a second device are presented in Supplementary Section~S1. As the micro-membrane has the STO layer, with high dielectric constant, facing upwards, the top-gate provides strong local coupling to the interfacial 2DEG. At positive $V_\mathrm{G}=0.7$~V the device exhibits high conductance, whereas decreasing $V_\mathrm{G}$ progressively depletes the 2DEG beneath the gate, ultimately leading to pinch-off of the conducting channel below a threshold voltage $V_\mathrm{th}$ (inset of Fig.~1e). 
This field-effect behavior is consistent with previous studies in bulk LAO/STO, where electrostatic gating has been employed to tune carrier density and drive metal-insulator transitions\cite{thiel2006tunable,caviglia2008electric}. Due to the much larger separation and smaller dielectric constants, the back-gate coupling to the 2DEG is considerably weaker. Only small changes of the threshold voltage are observed upon changing $V_\mathrm{BG} \pm 120 \, \mathrm V$. \\
\indent In addition to its stronger capacitive coupling, the lithographically defined local-gate provides spatially selective control that cannot be achieved with a global gate. In bulk LAO/STO, comparable local electrostatic control has been realized using both conductive-AFM patterning\cite{cen2009oxide, cheng2011sketched, cheng2015electron, tomczyk2016micrometer, annadi2018quantized}  and lithographically defined metallic gates\cite{goswami2015nanoscale, prawiroatmodjo2017transport}, enabling the formation of nanowires, tunnel barriers, and QDs.

\subsection*{Negative-$U$ Double Quantum Dot}
\indent The results from the single-gate devices demonstrate strong local electrostatic control of the carrier density in transferred freestanding micro-membranes, enabling complete pinch-off using relatively small gate voltages due to the efficient top-gate/2DEG coupling through the STO layer in devices with the orientation flipped compared to the growth direction. Such control is key for realizing tunable confinement in semiconductor quantum devices. To investigate this regime in the transferred membranes we fabricated a double local-gate device featuring a narrow constriction controlled by two local-gates, as shown in Fig.~2a. Transport measurements were performed in a dilution refrigerator at a base temperature $T_\text{base}\approx$ 20 mK under an out-of-plane magnetic field $B_\perp=200$ mT, sufficient to suppress superconductivity of the LAO/STO 2DEG\cite{erlandsen2022two} while preserving the negative-$U$ character\cite{cheng2015electron, meucci2026intrinsic}. \\
\indent Figure 2c shows the differential conductance, $G$ at zero source-drain bias, $V_{\text{SD}}=0$, as a function of the two local-gate potentials, $V_{\text{G}1}$ and $V_{\text{G}2}$, measured at a fixed back-gate potential $V_{\text{BG}}=-7$ V. For large positive values of either $V_{\text{G}1}$ or $V_{\text{G}2}$, the device is highly conductive (regimes I and II), corresponding to the situations illustrated schematically in Fig.~2b, where transport occurs through the regions beneath the gates. As the gate potentials are reduced, the 2DEG beneath the gates is progressively depleted, and transport becomes confined to the narrow constriction between them (regime III). In this regime, a series of conductance resonances emerges. Varying $V_{\text{BG}}$ preserves the overall pattern while shifting the  resonance positions in $(V_{\text{G}1},V_{\text{G}2})$  space (Supplementary Section~S2).\\
\indent Figure 2d shows $G$ as a function of $V_{\text{SD}}$, measured along the white dashed line in Fig.~2c by sweeping $V_{\text{G}1}$ while keeping $V_{\text{G}2}=-0.5$ V.  The characteristic diamond pattern shows that the resonances observed in Fig.~2c arise from transport through a QD in Coulomb-blockade. Similar features are observed for cuts along other directions in $(V_{\text{G}1},V_{\text{G}2})$  space (Supplementary Section~S3).\\
\indent The Coulomb peaks in Fig.~2c can be grouped into two families according to their slopes (red and yellow arrows), indicating that transport is dominated by two separate QDs, QD1 and QD2, with different electrostatic coupling to the two top-gates.\\
\indent The formation of QDs is common in gated nano-devices near pinch-off, where spatial potential fluctuations arising from impurities, surface charges, and other disorder sources can localize charge carriers. Such effects enabled early studies of QDs in semiconductor nanowires\cite{weis2014quantum}, silicon heterostructures\cite{scott1989conductance}, and gated LAO/STO devices\cite{prawiroatmodjo2017transport}. Similarly, we attribute the formation of the QDs observed here to potential fluctuations that break the gate-induced one-dimensional confinement into separate local minima near pinch-off.\\
\indent Figure 2f shows $G(V_\mathrm{G1})$ for a fixed $V_\mathrm{G2} = -0.5 \, \mathrm V$ and increasing in-plane magnetic field, $B_\parallel$. The Coulomb resonances bifurcate above a characteristic pairing field, $B_\mathrm{P}$, with a splitting that increases linearly with $B$. Consistently, bias spectroscopy reveals a doubling of the Coulomb-diamond structure above $B_\text{P}$ (Fig.~2e). These observations are fully consistent with previous reports of electron pairing and negative-$U$ physics in LAO/STO QDs\cite{cheng2015electron, prawiroatmodjo2017transport, meucci2026intrinsic}. Notably, the pairing persists above both the superconducting transition temperature and the critical magnetic field (Supplementary Section~S4). \\
\indent Figure 3a shows a high-resolution zoom of Fig.~2c. The Coulomb resonances exhibit clear avoided crossings arising from interdot coupling, giving rise to the honeycomb pattern characteristic of the charge-stability diagram of a serially coupled DQD.  Because of the negative $U$, the ground-state charge configurations, $(Q_1,Q_2)$, of the two QDs correspond to states $(eN_1,eN_2)$, where $N_1$ and $N_2$ are even positive integers. Since the absolute charge occupancies cannot be determined experimentally, we label all charge states in Fig.~3a relative to unknown even charge offsets, $(eN_1^0,eN_2^0)$.  \\
\indent 

\subsection*{Characteristic Energy Scales}
\indent The characteristic energy scales governing the separate QDs are found by adapting the  analysis of negative-$U$ QDs\cite{cheng2015electron} in regions away from the avoided crossings. From bias spectroscopy (Fig.~2d), the energy associated with the addition of two electrons to a single QD is extracted as $E_\text{add}=350\pm50$ \textmu eV, and the lever-arms, $\alpha_{ij}$, characterizing the coupling of QD $i$ to top-gate $j$ are found from the slopes of the Coulomb-blockade diamonds (Supplementary Section~S5). The lever arm for coupling to gate $1(2)$ is independent of $V_{\mathrm G2} (V_\mathrm{G1})$, and their relative magnitudes show that QD1 is tuned approximately equally by both gates, while QD2 is mainly controlled by $V_\mathrm{G2}$. This is consistent with the overall resonance slopes in $(V_\mathrm{G1},V_\mathrm{G2})$ space (Figs.~2c and 3a) and suggests a geometry in which QD1 is positioned approximately symmetrically between the two gates while QD2 is situated closer to gate 2.\\
\indent To extract the attractive on-site interaction $|U|$, the field-induced peak splitting $\Delta V_{\text{G}j}(B)$ of resonances associated with QD$i$ is found by fitting $G (V_{\text{G}j})$ to a double Lorentzian (Fig.~2f). The pairing field, $B_{\mathrm{P}}$, is obtained by extrapolating $\Delta V_{\mathrm{G}j}(B)$ to zero, where $B=\sqrt{B_\parallel^2+B_\perp^2}$.  The $g$-factor is obtained from the slope of the linear field dependence as $\mu_B g_{ij} = e \alpha_{ij} d \Delta V_{\text{G}j}(B)/dB$ (Supplementary Section~S5). Within fitting uncertainties, $B_\text{P}=0.4\pm0.1$ T and $g=1.5\pm0.1$ do not vary among different resonances of the same QD, between the two QDs, or with $V_{\text{G}1}$ and $V_{\text{G}2}$ (Supplementary Section~S5). Finally, the on-site interaction is estimated as $|U|=g\mu_\mathrm{B} B_\mathrm{P} = 35 \pm 8 \mathrm{ \mu eV}$.  \\
\indent Compared to a single QD, a DQD introduces two additional parameters: the interdot tunneling amplitude, $t$, and the interdot Coulomb interaction, $W$. Figure 3c illustrates the charge-stability diagrams expected in the negative-$U$ regime for different values of $t$, $W$, and $B$.   In conventional positive-$U$ DQDs, the interdot parameters are estimated from the charge-stability diagram, where $W$ is positive due to Coulomb repulsion and determines the rigid splitting of the triple points, while $t$ introduces avoided crossings between coupled states. This situation is illustrated schematically in Fig.~3c, panels $v$, $vi$, $viii$ and $ix$. Since the intradot effective interactions are attractive $(U<0)$ we also allow for the case of attractive interdot coupling, corresponding to $W<0$ (Fig.~3c, panels $i$, $iv$, $vii$, $x$). In this regime, the tunneling-induced splitting of the triple points can be compensated by $W$, and for sufficiently large $|W|$ the qualitative pattern of triple-point charge degeneracies changes (Fig.~3c, panels $iv$ and $vii$). Furthermore, the LAO/STO 2DEG may exhibit significant spin-orbit coupling\cite{caviglia2010tunable, lesne2016highly}, giving rise to a spin-flipping tunneling amplitude, $t_\mathrm{SO}$, which also modifies the charge-stability diagram (Fig.~3c, panels $x$-$xii$). The parameters $W$, $t$, and $t_\mathrm{SO}$ can be extracted from the magnitude of the avoided crossings in the stability diagram at $B_\parallel=0$ and $B>B_\mathrm{P}$. At zero field, $\Delta_{02-20}=2W-\Lambda+\sqrt{\smash[b]{\Lambda^2+4t_{\mathrm{eff}}^2}}$, where $\Lambda=(|U|+W)/2$ and $t_{\mathrm{eff}}={\sqrt{\smash[b]{t^2+t_{\mathrm{SO}}^2}}}$ (Fig.~3b, Supplementary Section~S6). \\
\indent  For $B>B_\mathrm p$ the conductance resonances split, allowing the odd-$N$ ground states (Fig.~2f). Figure 3g shows the experimental evolution of the stability diagram with $B_\parallel$, showing the doubling of the honeycomb pattern as the odd-$N$ states are stabilized. Figure 3d shows a zoom of the $B_\parallel = 3\,\mathrm T$ measurement where the $(1,1)$ charge state is clearly resolved. The expected splittings of the spectrum $\Delta_{01 - 10}^{B>B_\mathrm{P}} = W+2t$ and $\Delta_{02-11}^{B>B_\mathrm{P}} = W+2\sqrt{2}t_{\mathrm{SO}}$  are indicated in yellow in Fig.~3e (Supplementary Section~S6). Notice that at $B>B_\mathrm{P}$ the $(0,2)$ and $(1,1)$ ground states are a singlet (S) and triplet (T), respectively, and the splitting $\Delta_{02 - 11}^{B>B_\mathrm{P}}$ therefore does not depend on $t$. The values for $W\approx0$ and $t_\mathrm{SO}\approx0$ can be inferred from the absence of a splitting at the $(0,2)-(1,1)$ charge degeneracy in Fig.~3d. To this end, Fig.~3f, shows line profiles extracted along the detuning axis, $\delta$, the energy axis, $\epsilon$, and the axis $\epsilon_\mathrm{S}$ parallel to $\epsilon$ and passing through the  $(0,2)-(1,1)$ degeneracy point. The red trace shows Lorentzian fits. No splitting is observed at the $(0,2) -(1,1)$ degeneracy point along $\delta$ and $\epsilon_\mathrm{S}$. We consider two cases: if $W> 0$, a splitting of magnitude $\Delta_{02-11}^{B>B_\mathrm{P}}$ should appear along  $\epsilon_\mathrm{S}$. Since no such splitting can be resolved and the peak width at the degeneracy point is comparable to the widths of the individual conductance peaks measured away from the degeneracy,  we rule out the $W>0$ scenario and conclude that $W+2\sqrt{2}t_{\mathrm{SO}}\leq 0$ (Supplementary Section~S7). If $W<0$, the avoided crossing $\Delta_{02-11}^{B>B_\mathrm{P}}$, is instead expected to appear along $\delta$. Even in the special case $W=-2\sqrt{2}\,t_{\mathrm{SO}}$, where the splitting vanishes, a substantial broadening of the resonance is expected along $\delta$. No such splitting or broadening is observed experimentally and thus we also rule out the $W<0$ scenario and conclude that both $W$ and $t_{\mathrm{SO}}$ are negligible. However, given the measured peak widths at and away from the degeneracy point, the data do not permit a quantitative upper bound to be placed on either parameter (Supplementary Section~S7).\\
\indent Setting $W=t_{\mathrm{SO}}=0$, fitting the avoided crossing $\Delta_{01 - 10}^{B>B_\mathrm{P}}$ in Fig.~3d yields $t\approx80 ~\mu\mathrm{eV}$ in agreement with the value extracted from $\Delta_{12 - 21}^{B>B_\mathrm{P}}$, $t\approx90 ~\mu\mathrm{eV}$. Fitting the $B_\parallel=0$, $02 - 20$ charge-transition branches, shown by white dashed lines in Fig.~3a, yields  $t\approx60 \pm30~\mu\mathrm{eV}$, where the uncertainty correspond to the characteristic scale of the discrete shifts of data-points, attributed to effects of coupling to charge traps in the vicinity of the DQD (Supplementary Section~S5). The three separate estimates of $t$ are in reasonable agreement, and we adopt $t=80 \pm 10~\mu\mathrm{eV}$ in the following. The combined parameters characterizing the LAO/STO DQD are collected in Supplementary Section~S8. \\
\indent With the full set of experimentally extracted parameters, we model the negative-$U$ DQD with an Anderson Hamiltonian\cite{anderson1975model} and calculate transport using a Pauli master-equation approach (Supplementary Section~S9)\cite{KIRSANSKAS2017317}. For each of the stability diagrams in Fig.~3a,d,g the corresponding calculated  results are shown in panels b,e, and h. In all cases the calculations are in excellent agreement with the experimental observations.

\begin{figure}[hbt]
\centering
\includegraphics[width = \linewidth]{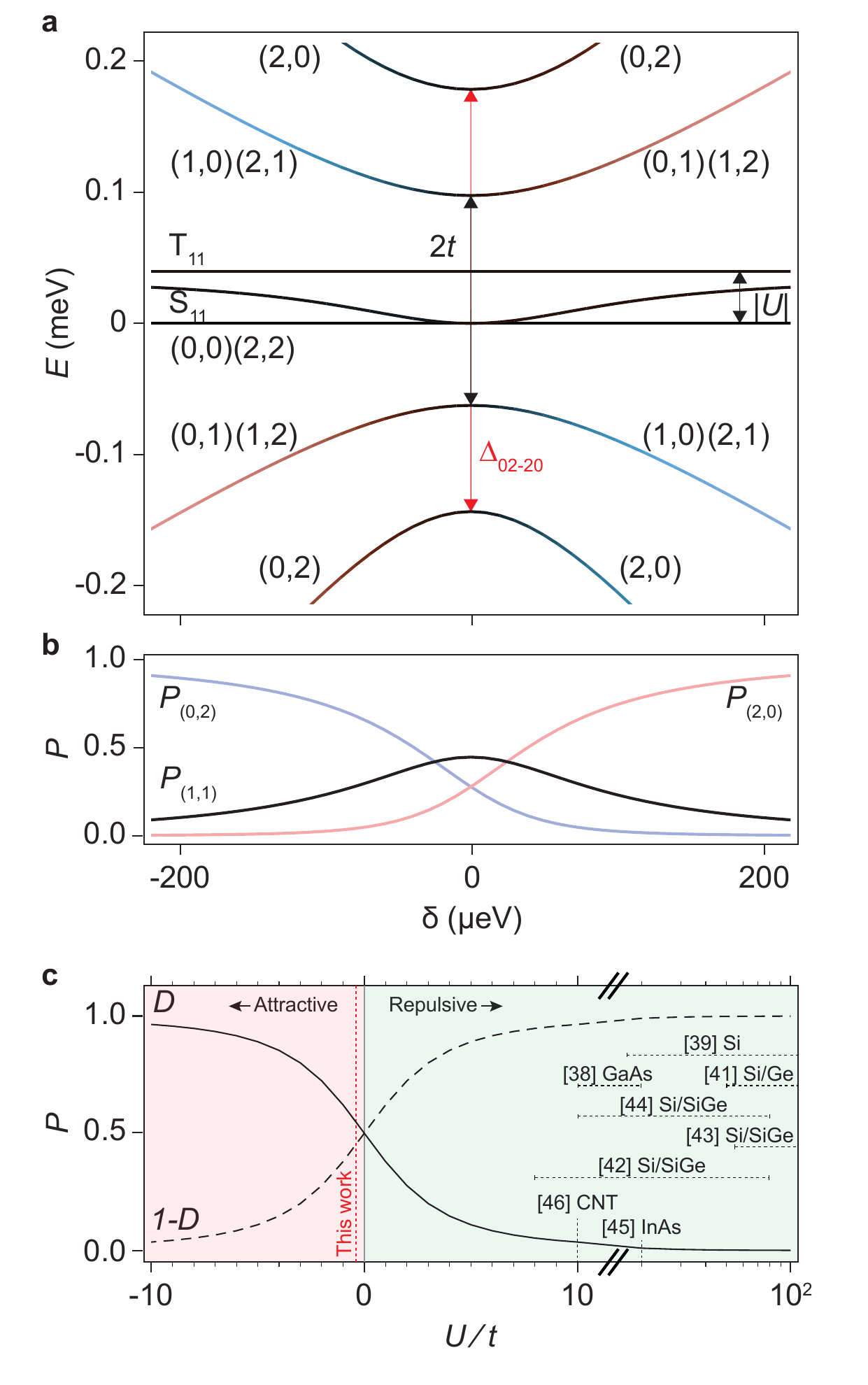}
\caption{ \textbf{Energy spectrum and paired ground state.}
\textbf{a} Energy levels of the negative-$U$ DQD as a function of detuning, $\delta$, calculated using experimentally extracted parameters. Line colors indicate the relative QD occupation: red denotes greater occupation of QD2, blue greater occupation of QD1, and black equal occupation of both QDs.The charge states $(N_1,N_2)$ are relative to unknown even offsets.
\textbf{b} Ground-state occupations $P_{02}$, $P_{20}$, and $P_{11}$ as a function of detuning, $\delta$. The combined occupation of the paired states, $P_{02}+P_{20}$, always exceeds $P_{11}$.
\textbf{c} Probability of double occupancy, $D$, and single occupancy, $1-D$, in a half-filled two-site Hubbard model as a function of $U/t$. At $\delta=0$ $D=P_{02}+P_{20}$ and $1-D=P_{11}$. The negative-$U$ DQDs (red), including the one realized here (red dashed line), are dominated by paired configurations, whereas positive-$U$ DQDs (green) are dominated by the $(1,1)$ state. Dashed lines mark values/ranges of $U/t$ reported in the literature. Note that the horizontal axis is linear between $-10<U/t<10$ and logarithmic above this range.
}
\end{figure}

\section*{DISCUSSION}
\indent DQDs constitute a canonical platform for quantum electronics and form the basis of charge, spin, and spin-orbit qubits in devices realized in semiconductor heterostructures, semiconductor nanowires, carbon nanotubes, and 2D materials\cite{van2002electron}. The device studied here represents the first DQD realized in an oxide host and the first realization of a DQD in the negative-$U$ regime, where on-site pairing is energetically favored. As a result, it occupies a regime that is qualitatively distinct from that of conventional semiconductor DQDs. \\
\indent Figure 4a shows the calculated energy spectrum as a function of detuning, $\delta$, which drives charge transfer between the two QDs. In contrast to conventional positive-$U$ DQDs (Supplementary Section~S10), the paired charge states $(2,0)$ and $(0,2)$ are energetically stabilized relative to the singly occupied $(1,1)$ configuration. \\
\indent The zero-field stability diagram for $U<0$ (Fig.~3a) qualitatively resembles that of a conventional $U>0$ DQD, except that transport is dominated by charge transfer in units of $2e$ electron pairs. In a conventional one-electron DQD, however,  the $(0,1)$ and $(1,0)$ states couple directly through the single-electron tunnel coupling $t$, forming bonding and anti-bonding states with equal weight at zero detuning approximately independent of $U/t$. In contrast, the paired $(0,2)$ and $(2,0)$ states of the negative-$U$ DQD are coupled through virtual occupation of the intermediate $(1,1)$ configuration. As a consequence, the ground state for $t>0$ contains a finite $(1,1)$ weight reflecting the competition between tunneling and pairing. This is illustrated in Fig.~4b, which shows the ground-state composition along the detuning axis where the hybridized ground state retains a substantial $(1,1)$ character at $\delta = 0$. Nevertheless, the combined weight of the paired configurations, $P_{02}+P_{20}$, exceeds that of the $(1,1)$ state, $P_{11}$, consistent with the pair dominated ground state.\\
\indent As confirmed by the calculations in Fig.~3, the observed DQD is well-described by an attractive two-site Hubbard model. To relate the present device regime to the repulsive case in conventional DQDs, Fig.~4c shows the double-occupancy probability $D = \langle n_{1\uparrow}n_{1\downarrow}+ n_{2\uparrow}n_{2\downarrow} \rangle$ of the two-site Hubbard model at half filling as a function of $U/t$. At the left-right symmetric point in gate space ($\delta=0$), $D$ corresponds to $P_{02}+P_{20}$, while $P_{11}=1-D$ gives the probability of single occupancy. Reversing the sign of $U$ changes the dominant character of the DQD ground state: whereas positive-$U$ DQDs are dominated by $(1,1)$ occupation, negative-$U$ DQDs favor paired configurations. In GaAs\cite{dicarlo2004differential}, Si\cite{lai2011pauli, das2011hubbard}, SiGe\cite{hu2007ge}, Si/SiGe\cite{das2011hubbard, simmons2009charge, zajac2015reconfigurable, wang2024efficient}, InAs\cite{pfund2006top}, and carbon nanotube\cite{ingerslev2008singlet} DQDs, typical values of $U/t$ range from $\sim5$--10 in strongly coupled devices and diverge in weakly coupled devices where $t\ll U$ (Fig.~4c, Supplementary Section~S8). For such large values of $U/t$, $1-D\approx1$ and the low-energy states are well approximated by the localized $(1,1)$ charge configuration. By contrast, the comparatively small value $|U|/t\approx0.4\pm0.1$ realized here places the device in a regime of substantial hybridization, and the resulting ground state is neither a localized electron pair nor a conventional singly occupied DQD state. \\
\indent Another notable feature of the observed DQD is the combination of substantial interdot tunneling, indicating close proximity and significant orbital overlap between the two QDs, with negligible interdot Coulomb interaction. Conventional semiconductor DQDs typically exhibit $W/U=0.2-0.8$ (Supplementary Section~S8)\cite{hensgens2017quantum, lai2011pauli, das2011hubbard, simmons2009charge, wang2024efficient, pfund2006top}, while our measurements are consistent with $W=0$ despite $t\approx80~\mu$eV. A possible explanation is provided by the large dielectric constant of STO, which can strongly suppress long-range Coulomb interactions. In addition, we speculate that the same, as-yet-unknown mechanism responsible for the attractive on-site interaction may also suppress the effective interdot interaction, resulting in $W \approx 0$.\\
\indent Additionally, although strong spin-orbit coupling is generally reported in bulk LAO/STO systems\cite{caviglia2010tunable,lesne2016highly}, several mechanisms may contribute to the observed suppression of $t_{\mathrm{SO}}\ll t$ and the absence of a resolvable S-T anticrossing in Fig.~3d. First, the spin-orbit interaction in LAO/STO depends strongly on carrier density and orbital occupancy, with the largest Rashba coupling occurring once the $d_{xz}/d_{yz}$ bands are occupied to a particular level, whereas the single-band $d_{xy}$ regime for low carrier densities exhibits substantially weaker spin-orbit coupling\cite{joshua2012universal,zhong2013theory,caviglia2010tunable,ben2010tuning, vaz2020determining}. While the carrier density is not measured directly, the device is operated close to depletion, suggesting transport predominantly through $d_{xy}$-derived states. Second, the magnitude of the S-T avoided crossing depends on the orientation of the magnetic field relative to the effective spin-orbit field. Since the DQD forms in a disorder-defined potential close to depletion, the spin-orbit axis is not known a priori and may also be influenced by the local electrostatic environment created by the gates. A large projection of the spin-orbit field onto $B_{\parallel}$ would suppress the S-T anticrossing. Finally, $t_{\mathrm{SO}}$ depends on the characteristic size of the system, as previously discussed  for semiconductor DQDs\cite{froning2021strong}. Specifically, $t_{\mathrm{SO}}/t \sim a/l_{\mathrm{SO}}$, where $a$ is the interdot tunnel distance and $l_{\mathrm{SO}}=\hbar^2/(m^*\alpha_{\mathrm R})$ is the spin-orbit length, with $m^* \sim 1.3 m_e$ denoting the effective mass and $\alpha_\mathrm R$ the Rashba coefficient\cite{nadj2010spin}. At carrier densities corresponding to sole  $d_{xy}$ occupation, $\alpha_{\mathrm R} \lesssim \SI{30}{\milli\electronvolt\angstrom}$ \cite{vaz2020determining}, yielding  $l_\mathrm{SO} \gtrsim 20 \, \mathrm{nm}$. Consequently, spin-flip tunneling is expected to be weak if the interdot distance, $a$, is significantly smaller than this length scale. Further measurements of the role of the magnetic field orientation on the S-T crossing may clarify the role of spin-orbit coupling in mesoscopic LAO/STO devices. Additionally, studies in devices with deterministic control of the quantum-dot size and interdot spacing may help elucidate the origin of both the suppressed interdot interaction and the weak spin-orbit-assisted tunneling.\\
\indent  Beyond their fundamental significance, the paired states realized here may provide a foundation for new forms of charge-based quantum devices. In analogy with conventional charge qubits, coherent manipulation could in principle be performed within the subspace spanned by the paired charge states $(2,0)$ and $(0,2)$. Such an implementation would differ fundamentally from superconducting DQDs and Andreev molecules, where paired states arise through coupling to a superconducting reservoir\cite{de2010hybrid}. In a negative-$U$ DQD, the paired states are intrinsic to the QDs themselves and therefore form a discrete spectrum without an associated quasiparticle continuum above a superconducting gap. However, in this case, the strong intrinsic $(1,1)$ mixing plays an important role. Increasing $|U|/t$ increases the purity of the paired states, whereas state manipulation requires interdot coupling, which is mediated through the intermediate $(1,1)$ configuration.  Thus, some degree of $(1,1)$ mixing is unavoidable. Future devices with independent control of $U/t$ could  explore the balance between pair-state purity and coherent coupling.\\
\indent More generally, full electrostatic control of the DQD parameters would enable access to coupled systems with attractive ($U_{1,2}<0$), repulsive ($U_{1,2}>0$), or mixed-sign ($U_1<0$, $U_2>0$) interactions across the two QDs. Previous studies have shown that the effective interaction in LAO/STO QDs can be tuned via gate voltage and/or QD occupancy and even driven across the positive- to negative-$U$ crossover\cite{cheng2016tunable}. Whether similar control can be achieved for the interdot interaction $W$ remains an open question. If so, attractive interdot interactions ($W<0$) could stabilize paired states delocalized across multiple QDs\cite{hamo2016electron}, a qualitatively different regime from the one realized here, where $W=0$ is consistent with strongly localized electron pairs. Unlike on-site negative-$U$ pairing, which competes with the Zeeman energy and can be suppressed by magnetic field, pairing driven by $W<0$ does not directly incur a Zeeman energy cost and could therefore remain robust even at high magnetic fields. \\
\indent  More broadly, the local electrostatic control of transferred oxide membranes demonstrated here establishes a promising route toward semiconductor-integrated oxide nanoelectronics. Unlike conventional LAO/STO nanostructures, in which electrostatic gates are typically fabricated above the LAO layer, transferred membranes offer the flexibility to pattern gates on the high-dielectric-constant STO side, close to the interface, or to integrate the membranes onto substrates with pre-patterned buried-gate architectures, enabling device geometries similar to those used in state-of-the-art semiconductor QD platforms.  Additionally, membrane transfer provides a pathway for combining oxide-specific quantum functionalities, including attractive interactions, gate-tunable superconductivity\cite{thiel2006tunable, caviglia2008electric, erlandsen2022two, reyren2007superconducting}, ferroic order\cite{christensen2019strain,lu2019freestanding}, and large electro-optic responses\cite{anderson2025quantum}, with mature semiconductor technologies such as charge sensors, microwave resonators, integrated photonic circuits, and CMOS control electronics. Importantly, previous demonstrations of large ordered arrays of oxide micromembranes\cite{dahm2021size}, together with the compatibility of freestanding membranes with polymer-assisted transfer techniques, provide a pathway toward deterministic large-area integration of oxide quantum devices on semiconductor platforms, suggesting a route from individual transferred devices toward scalable architectures of many coupled oxide devices. \\
\indent  In conclusion, we have demonstrated local electrostatic control of freestanding LAO/STO micro-membranes transferred onto silicon substrates and  observed an oxide DQD operating in the negative-$U$ regime. This work realizes a previously unexplored class of DQD, combining two features that have not been reported before: a DQD formed in an oxide material and DQD operation in the negative-$U$ regime. The resulting device occupies an unusual region of DQD parameter space characterized by attractive on-site interactions, strong interdot hybridization, negligible interdot Coulomb coupling and negligible spin-orbit coupling. These results establish that electron-pairing physics survives membrane spalling and show that DQDs arising in transferred oxide membranes can be quantitatively characterized through transport spectroscopy and modeling. Although deterministic electrostatic definition of QDs remains to be demonstrated, the present results suggest that oxide membranes provide a viable platform for controlled engineering of quantum confinement in correlated-electron systems and open a pathway toward semiconductor-integrated quantum devices that exploit emergent functionalities unique to complex oxides.

\section*{MATERIALS AND METHODS}
\subsection*{Device fabrication}
Freestanding LAO/STO (LAO/STO thickness: 55 nm / 55 nm)  micromembranes were fabricated following the controlled spalling technique described in Ref.\ \citenum{sambri2020self, guarino2026growth}. The resulting membranes, with lateral dimensions
$\sim2\times2$ \textmu m,  were transferred onto a SiO$_2$ (thickness: 500 nm) / p$^{++}$ doped Si substrate. All data shown in the text are from devices having the LAO layer facing the substrate. 
Electrical contacts to the interfacial 2DEG were defined by electron-beam lithography and argon ion Kaufmann milling at $\sim50$° with respect to the membrane plane, followed by in-situ Ti/Au (15 nm / 600 nm) evaporation at $\sim70$° from two antiparallel in-plane orientations. The milling step and angled metal deposition are required to establish electrical contact to the buried 2DEG at the LAO/STO interface\cite{erlandsen2022two}. Single local-gates of varying width 100, 200 and 400 nm were defined by electron-beam lithography, followed by Ti/Au (5 nm / 100 nm) top-down evaporation. Double local-gates were defined by electron-beam lithography and SiO$_2$ evaporation, first at 90° (5 nm) and then at 60° (20 nm) relative to the membrane plane, followed by Ti/Au (5 nm / 130 nm) top-down evaporation.
\subsection*{Transmission electron microscopy}
A cross-sectional transmission electron microscopy (TEM) lamella was extracted from the STO/LAO membrane by using a dual beam focus ion beam (FIB) - scanning transmission electron microscope (SEM) Thermo Fisher Helious 5UX. The sample was analyzed using aberration corrected high-angle annular dark field (HAADF) scanning TEM (STEM), on a double-corrected and monochromated ThermoFisher Spectra 300 microscope operating at 300 kV. STEM images were processed using Velox Software (ThermoFisher Scientific) and DigitalMicrograph (GATAN). Four dimensional (4D) STEM analysis were performed in a Gatan K3 direct electron detector coupled to a Gatan Image Filter Continuum in the same Spectra 300 TEM.
\subsection*{Transport measurements}
Room-temperature transport measurements were performed using standard direct-current techniques. Low temperature  measurements were carried out in a dilution refrigerator with base temperature $T_\text{base}=20$ mK under a small out-of-plane magnetic field $B_\perp=200$ mT, sufficient to suppress superconductivity in the membrane while preserving pair-mediated transport. Note that the field is out of plane relative to the substrate orientation. The differential conductance was measured using standard lock‑in techniques. \\
\subsection*{Transport simulations}
Details are provided in Supplementary Section~S9. 
\subsection*{Data analysis}
Details of the data analysis are provided in the main text and in the Supplementary Information. Where relevant, reported parameter values correspond to error‑weighted averages, and the associated uncertainties denote 68\% confidence intervals.

\nocite{ihn2009semiconductor}
\bibliography{ref}

\subsection*{Acknowledgments}
The authors acknowledge helpful discussions with Martin Leijnse and Jens Paaske.

\noindent \textbf{Funding:}

This work was supported by the European Research Council through the European Union’s Horizon 2020 research and innovation program (Grant No. 866158) and through the Horizon Europe EIC Pathfinder program Project IQARO (Grant No. 101115190).
The authors acknowledge support from FOXES - Freestanding Oxides Epitaxial Strained Micro-Heterostructures (PRIN PNRR 2022, Project No. 2022TCT72).
F.T. acknowledges support by research grant no. 37338 (SANSIT) from Villum Fonden and grant no. 10.46540/4254-00012B (PESTO) from Independent Research Fund Denmark.
ICN2 is supported by the Severo Ochoa program from Spanish MCIN / AEI (Grant No.: CEX2021-001214-S) and is funded by the CERCA Programme / Generalitat de Catalunya. ICN2 also acknowledges funding from Generalitat de Catalunya (2025SGR00117) and from 
Grant IU16-014206 (METCAM-FIB) funded by the European Union through the European Regional Development Fund (ERDF), with the support of the Ministry of Research and Universities, Generalitat de Catalunya.
We acknowledge support from CSIC Interdisciplinary Thematic Platform (PTI+) on Quantum Technologies (PTI-QTEP+). This research work has been funded by the European Commission – NextGenerationEU (Regulation EU 2020/2094), through CSIC's Quantum Technologies Platform (QTEP). Funded by the Horizon Europe call HORIZON-INFRA-2023-SERV-01 project RIANA under grant agreement number 101130652. Authors acknowledge the use of instrumentation as well as the technical advice provided by the Joint Electron Microscopy Center at ALBA (JEMCA).  

\noindent \textbf{Author contributions:}
Conceptualization: G.M., T.S.J.
Data curation: G.M, T.J., P.O.K., D.O., D.J.C.
Formal analysis: G.M., T.J.
Funding acquisition: T.S.J.
Investigation:  G.M, T.J., P.O.K., D.O., D.J.C., C.B.M.,  J.A., T.S.J.
Methodology: G.M., T.J., P.O.K., D.O., D.J.C.,  F.M.G., A.S, T.S.J.
Project administration: T.S.J.
Resources: G.M., T.J., P.O.K., D.O., D.J.C.,  F.M.G., A.S, T.S.J.
Supervision: T.S.J.
Validation: G.M, T.J., P.O.K., D.O., D.J.C., C.B.M., J.A., N.B., M.B., R.C., E.D.G., A.K., N.M., C.A.P., M.S., P.W., F.T., N.P., F.M.G., A.S, T.S.J.
Visualization: G.M., T.S.J.
Writing—original draft: G.M., T.S.J.
Writing—review and editing: G.M, T.J., P.O.K., D.O., D.J.C., C.B.M., J.A., N.B., M.B., R.C., E.D.G., A.K., N.M., C.A.P., M.S., P.W., F.T., N.P., F.M.G., A.S, T.S.J.

\noindent \textbf{Competing interests:}
The authors declare that they have no competing interests.

\noindent \textbf{Data, code, and materials availability:}
All data needed to evaluate the conclusions in the paper are present in the paper and/or the Supplementary Materials.

\end{document}


\preprint{AIP/123-QED}


\title{Supplementary Information for A Negative-$U$ Double Quantum Dot in a LaAlO$_3$/SrTiO$_3$ Membrane on Silicon}

\maketitle

\onecolumngrid

\section{Room-Temperature characterization of an additional single local-gate device}
\begin{figure}[H]
\centering
\includegraphics[width = \linewidth]{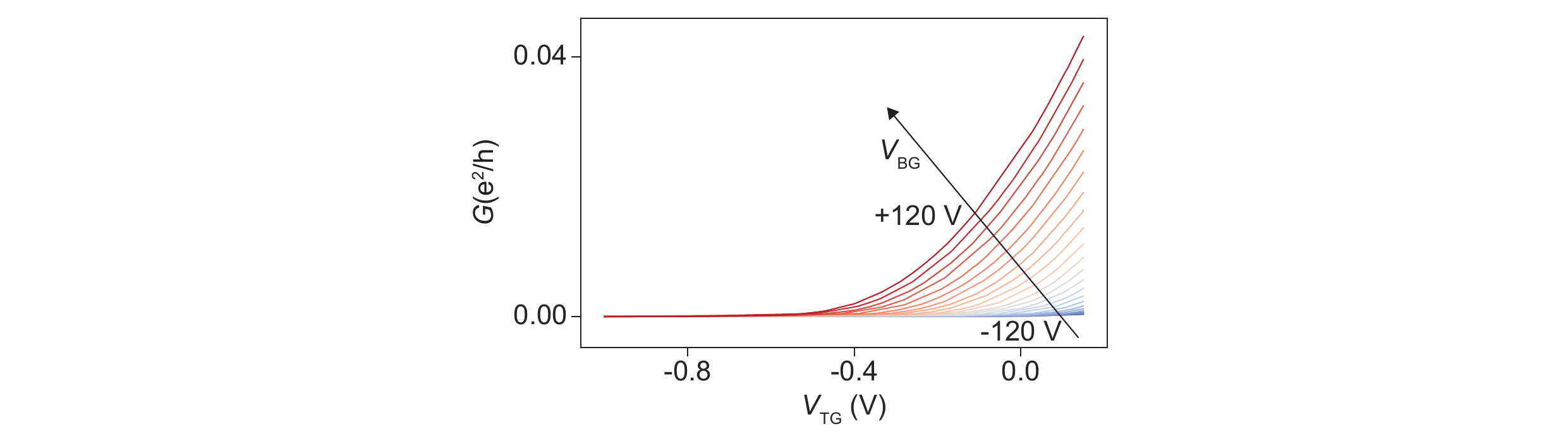}
\caption{ Room-temperature conductance, $G$, as a function of a single local-gate voltage, $V_\text{G}$, for back-gate voltages, $V_\text{BG}$, ranging from $-120$ to $+120$ V, measured in a device with top-gate width 400 nm. 
}
\label{S_FET2}
\end{figure}

Figure \ref{S_FET2} shows the room-temperature conductance, $G$, of a second device with top-gate width 400 nm, as a function of the local-gate potential, $V_\mathrm{G}$, for several values of the back-gate potential, $V_\mathrm{BG}$. The device exhibits behavior qualitatively similar to that of the 100 nm wide device discussed in the main text (Fig.~1e), including efficient local gating and modulation of the threshold voltage by the back gate. A comprehensive transistor characterization of these devices will be reported elsewhere.

\section{Global back-gate dependence}

\begin{figure}[H]
\centering
\includegraphics[width = \linewidth]{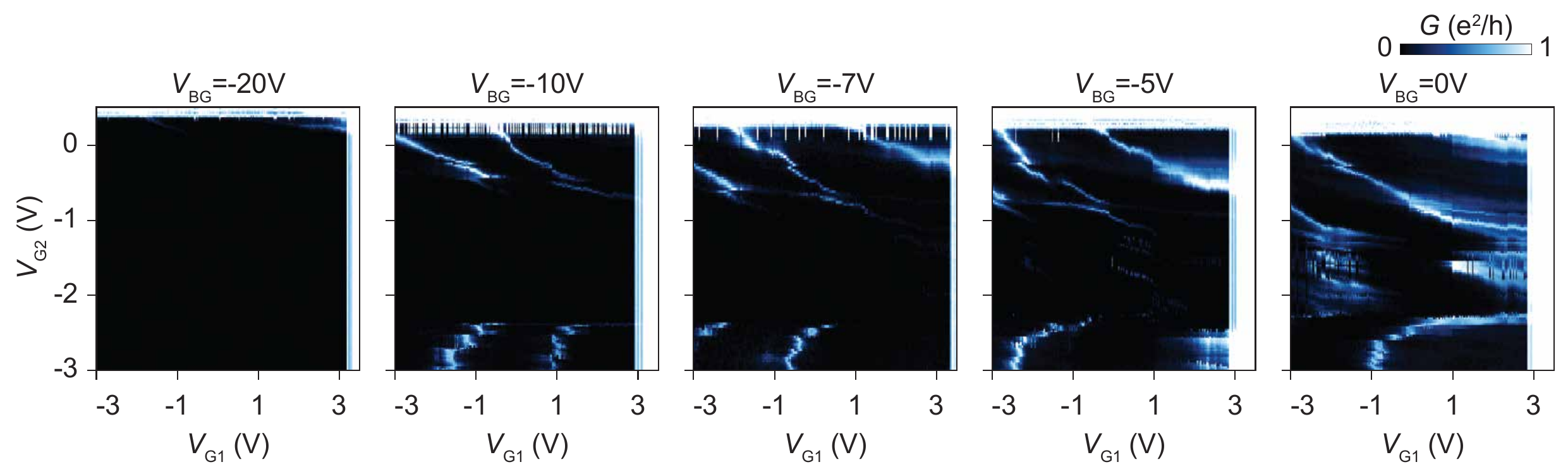}
\caption{ Evolution of the differential conductance as a function of local gate voltages for different values of the global gate voltage $V_\text{BG}$. Measurements performed at  $T_\text{base}=20$ mK, $B_\parallel=0$, $B_\perp=200$ mT.  
}
\label{S_backgate}
\end{figure}

The global back‑gate voltage shifts the threshold voltage of the local gates required to pinch off the channel, and the quantum‑dot resonances in gate‑voltage space. The resulting evolution of the charge‑stability diagram under back‑gate tuning is shown in Fig.\ \ref{S_backgate}.

\section{Bias spectroscopy}
\begin{figure}[H] 
\centering 
\includegraphics[width = \linewidth]{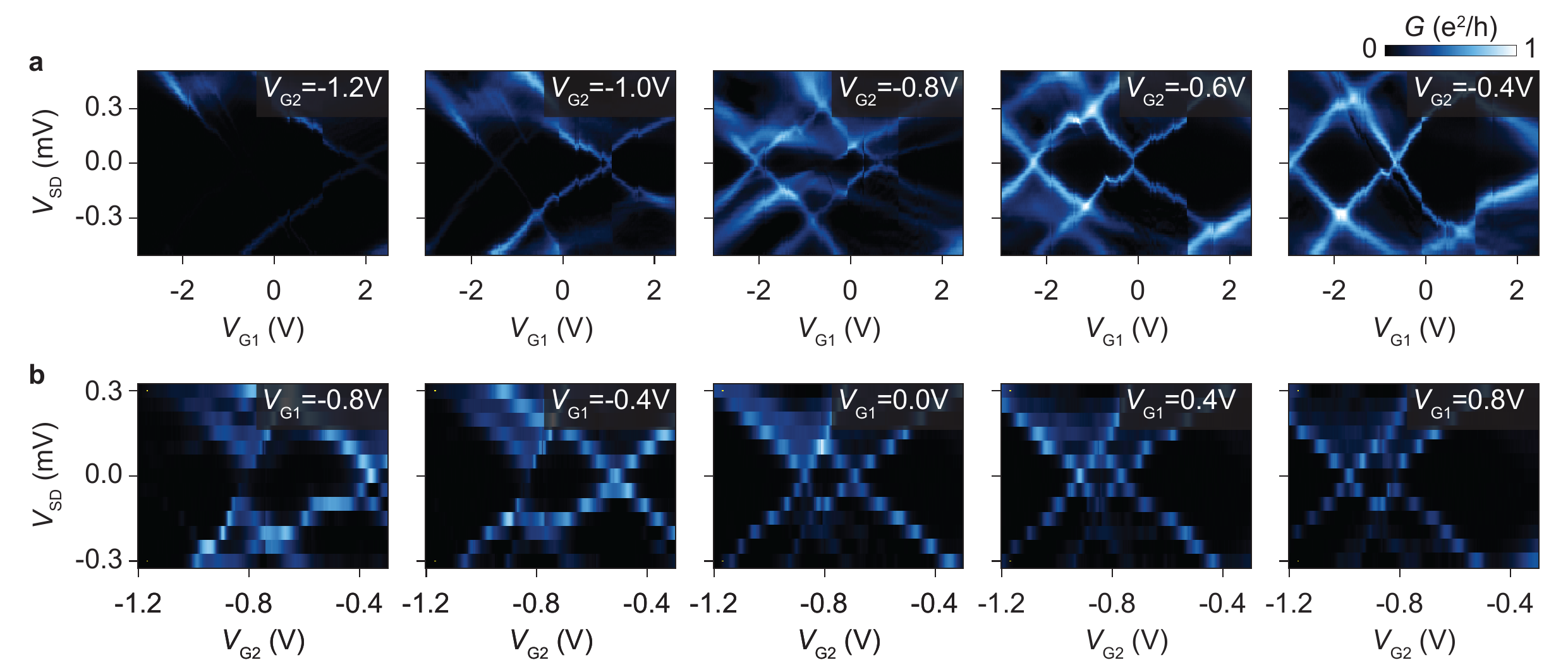}
\caption{ \textbf{a} Differential conductance as a function of $V_\mathrm{SD}$ measured while sweeping $V_{\mathrm{G}1}$ for several fixed values of $V_{\mathrm{G}2}$. \textbf{b} Differential conductance as a function of $V_\mathrm{SD}$ measured while sweeping $V_{\mathrm{G}2}$ for several fixed values of $V_{\mathrm{G}1}$. } 
\label{S_biasspec} 
\end{figure}

The differential conductance, $G$, was measured as a function of the source-drain bias, $V_\mathrm{SD}$, along different directions in $(V_{\mathrm{G}1},V_{\mathrm{G}2})$  space. Figure~\ref{S_biasspec}a presents line cuts along $V_{\mathrm{G}1}$ for different fixed values of $V_{\mathrm{G}2}$, while Fig.~\ref{S_biasspec}b shows line cuts along $V_{\mathrm{G}2}$ for different fixed values of $V_{\mathrm{G}1}$. 

\section{Negative-$U$ behaviour above the critical superconducting temperature}
\begin{figure}[H]
\centering
\includegraphics[width = \linewidth]{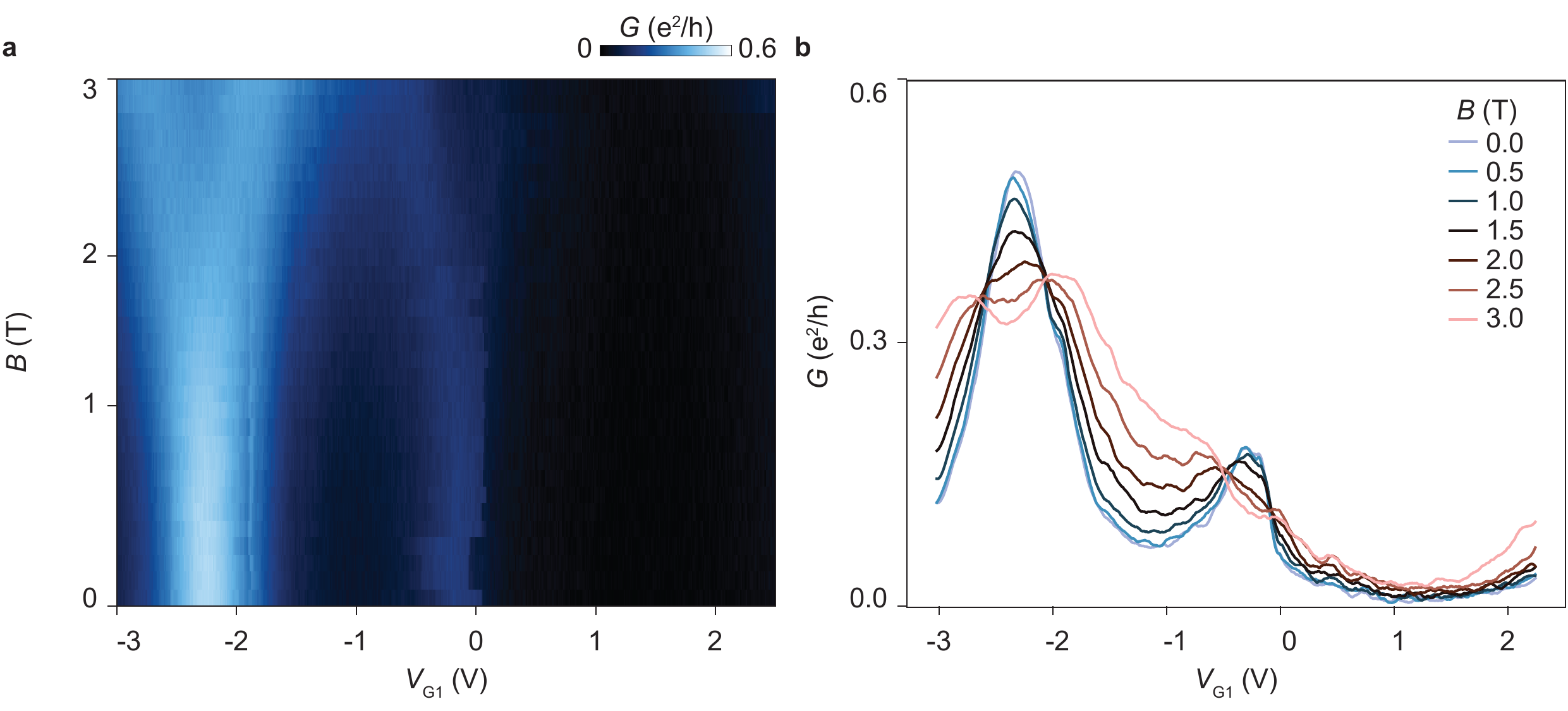}
\caption{ 
\textbf{a} Differential conductance as a function of  $V_{\text{G}1}$ and in-plane magnetic field $B_{\parallel}$, at $T=300$mK.
\textbf{b} Selected linecuts at $B_{\parallel} = 0$, 0.5, 1, 1.5, 2, 2.5, and 3 T showing the persistence of peak splitting with increasing magnetic field.
Measurements performed at $V_{\text{G}2}=-0.7$ V $V_\text{BG}=-7$V,   $B_\perp=200$ mT.  
}
\label{S_T=300mK}
\end{figure}

Negative‑$U$ pairing persists at temperatures and magnetic fields exceeding the critical values of superconductivity of the two dimensional electron gas. Fig.\ \ref{S_T=300mK} shows  that peak splitting associated with negative‑$U$ remains visible at $T=300$ mK and $B_{\perp}=200$, both above the superconducting critical temperature $T_\text{c}\approx220$ mK and critical field $B_{\perp\text{c}}\approx150$ mT. 

\section{Characteristic energy scales}
\begin{figure}[H]
\centering
\includegraphics[width = \linewidth]{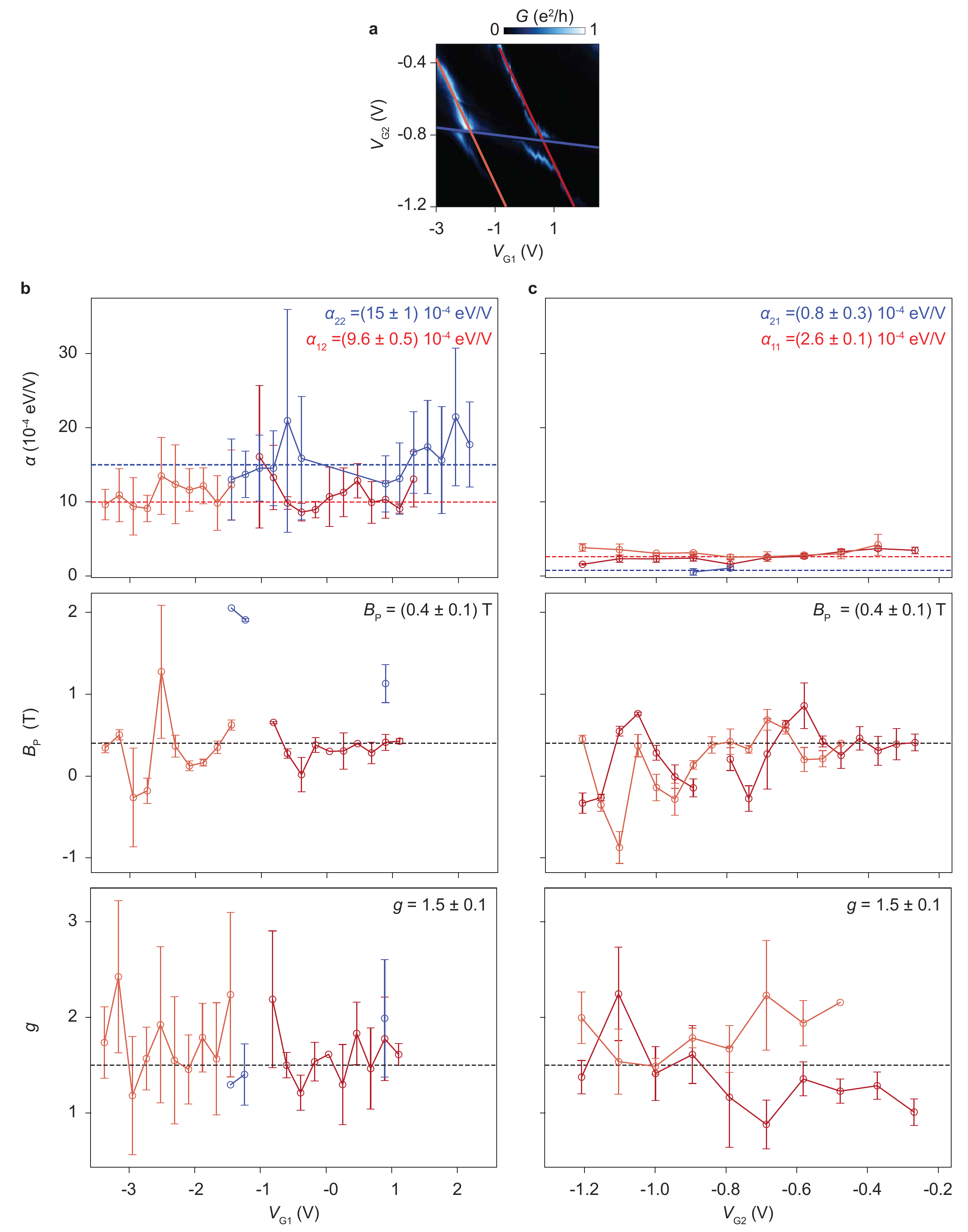}
\caption{
\textbf{a} Differential conductance as a function of $V_{\text{G}1}$ and $V_{\text{G}2}$ at $B_{\parallel}=0$. The lines indicate the three representative resonances used in the analysis. 
\textbf{b} Extracted $\alpha_{12}$, $\alpha_{22}$, $B_{\text{P}\parallel}$ and $g$, for the three resonances as a function of $V_{\text{G}1}$.
\textbf{c} Extracted $\alpha_{11}$, $\alpha_{21}$, $B_{\text{P}\parallel}$ and $g$, for the three resonances as a function of $V_{\text{G}2}$. The dashed lines in panel b and c indicate the error-weighted averages reported in the main text.
}
\label{S_gatedependence}
\end{figure}

The characteristic energy scales governing the separate quantum dots, including the gate lever arms, pairing energy, and effective $g$-factor, are extracted as functions of the two local gate voltages $V_{\text{G}1}$ and $V_{\text{G}2}$ for representative resonances of the two quantum dots. In particular, we focus on three resonances, two associated with one quantum dot and one with the other, as shown in Fig.\ \ref{S_gatedependence}. For each resonance of quantum dot $i$, we determine:

\begin{itemize}

    \item The lever arms with respect to the two gates, $\alpha_{i1}$ and $\alpha_{i2}$ as functions of the opposite gate voltage.  
    At fixed $V_{\text{G}1}$ ($V_{\text{G}2}$), $\alpha_{i2}$ ($\alpha_{i1}$) is extracted from bias spectroscopy measurements by fitting the four arms of the Coulomb resonance. Each arm provides an independent estimate of the lever arm, and the final value is obtained by averaging the four extracted values. The associated uncertainty is taken as the standard deviation of these four values. This procedure is repeated as a function of $V_{\text{G}1}$ ($V_{\text{G}2}$). 

    \item The pairing field $B_{\text{P}\parallel i}$ and the effective $g$-factor $g_i$ as functions of $V_{\text{G}1}$ ($V_{\text{G}2}$). To this end, we analyze conductance traces $G$ versus $V_{\text{G}2}$ ($V_{\text{G}1}$) measured at fixed $V_{\text{G}1}$ ($V_{\text{G}2}$) and varying magnetic field $B\parallel$. For each magnetic field, the resonance peaks are fitted with single, double, or triple Lorentzian functions, depending on the number of resolved features. In some cases, triple-Lorentzian fits are required to account for additional nearby resonances appearing close to the resonance of interest.

    From the fitted peak positions, we extract the splitting $\Delta V_{\text{G}2}$ ($\Delta V_{\text{G}1}$) as the separation between the relevant peaks at each magnetic field. The uncertainties are obtained from the covariance matrix of the fit and propagated accordingly. For each fixed value of $V_{\text{G}1}$ ($V_{\text{G}2}$), $\Delta V_{\text{G}2}$ ($\Delta V_{\text{G}1}$) is then plotted as a function of $B=\sqrt {B_\parallel^2+B_\perp^2}$, where $B_\perp=200$ mT, and fitted with a linear model over the high field range where the dependence is linear.

    The pairing field $B_{\text{P}i}$ is defined as the zero-crossing of this linear fit, and its uncertainty is obtained by standard error propagation from the fitted slope and intercept. 
    The effective $g$-factor is extracted from the slope $m$ of the linear fit according to
    \[
        m = \frac{\mu_B g_i}{e \alpha_{ij}},
    \]
    where $\mu_B$ is the Bohr magneton and $\alpha_{ij}$ is the corresponding gate lever arm.
\end{itemize} 
\indent We find that the lever arms $\alpha_{ij}$ depend on both the quantum dot index $i$ and the gate index $j$, but do not vary between different resonances of the same dot, nor do they depend on the opposite gate voltage. The magnitudes of the lever arms are $\alpha_{11}=(2.6\pm0.1)\cdot10^{-4}$, $\alpha_{12}=(9.6\pm0.5)\cdot10^{-4}$, $\alpha_{21}=(0.8\pm0.3)\cdot10^{-4}$, $\alpha_{22}=(15\pm1)\cdot10^{-4}$. The extracted pairing field $B_{\text{P}}$ and effective $g$-factor do not depend significantly on the quantum dot index, the specific resonance, or the local gates (see Fig.\ \ref{S_gatedependence} ). In the main text, we therefore report error-weighted averages over the different estimates and quote uncertainties corresponding to the $68\%$ confidence interval.\\
\indent The interdot tunneling amplitude, $t$, is first estimated from the magnitude of the $\Delta_{01-10}^{B>B_\mathrm{P}}$ avoided crossing in Fig.~3d. A quadruple-Lorentzian fit to the line profile along the energy axis, $\epsilon$, (Fig.~3f) yields $\Delta_{01-10}^{B>B_\mathrm{P}}=166\pm1$~\textmu eV, corresponding to $t=\Delta_{01-10}^{B>B_\mathrm{P}}/2=83\pm1$~\textmu eV. Similarly, fitting the $\Delta_{12-21}$ avoided crossing gives $t=89\pm1$~\textmu eV. \\
\indent The charge-transition branches in Fig.~3a are fitted to obtain additional estimates of $t$ and $\alpha_{ij}$. Assuming negligible interdot charging energy, $W$, and spin-orbit tunneling amplitude, $t_\mathrm{SO}$, as justified in the main text, the branches are fitted to the standard expressions \[ \mu_1=\epsilon-\frac{\sqrt{\delta^2+\Delta_{02-20}^2}}{2}=0, \qquad \mu_2=\epsilon+\frac{\sqrt{\delta^2+\Delta_{02-20}^2}}{2}=0, \] where $\epsilon=(\epsilon_1+\epsilon_2)/2$ is the average energy of the two quantum dots and $\delta=\epsilon_1-\epsilon_2$ is the interdot detuning [58]. The energy of quantum dot $i$ is given by $\epsilon_i=k_i-e\sum_j\alpha_{ij}V_{\mathrm{G}j}$, where $k_i$ is a constant offset defining the zero-gate reference level. The fit, shown by the white dashed lines in Fig.~3a, yields lever arms $\alpha_{11}=2.5\cdot10^{-4}$, $\alpha_{12}=7.5\cdot10^{-4}$, $\alpha_{21}=0.3\cdot10^{-4}$, $\alpha_{22}=12\cdot10^{-4}$, consistent with those extracted from bias spectroscopy and $\Delta_{02-20}=109$~\textmu eV. From this value we obtain $t=63\pm2$~\textmu eV. However, this uncertainty does not account for the discrete shifts of data points in the vicinity of $\Delta_{02-20}$, which correspond to a characteristic energy scale of approximately $t=30$~\textmu eV. Taking this additional source of uncertainty into account, a more realistic estimate is $t=60\pm30$~\textmu eV. 

\section{Energy splitting and avoided crossings}
To determine the characteristic energy splittings of the DQD, we first consider the spectrum of the uncoupled system, neglecting both spin-conserving tunneling, $t$, and spin-flipping tunneling, $t_{\mathrm{SO}}$. The energies of the charge configurations $(eN_1,eN_2)$ with $N_1+N_2\leq 4$ are 
\begin{align*} 
E_{00} &=0, \\ 
E_{10} &= \epsilon_1 ,\\ 
E_{01} &= \epsilon_2 ,\\ 
E_{11} &= \epsilon_1+\epsilon_2+W ,\\ 
E_{20} &= 2\epsilon_1+U ,\\ 
E_{02} &= 2\epsilon_2+U ,\\ 
E_{21} &= 2\epsilon_1+\epsilon_2+U+2W ,\\ 
E_{12} &= \epsilon_1+2\epsilon_2+U+2W ,\\ 
E_{22} &= 2\epsilon_1+2\epsilon_2+2U+4W . 
\end{align*} 
Here, $\epsilon_i$ is the single-particle energy of quantum dot $i$, $U$ is the on-site interaction energy, and $W$ is the interdot Coulomb interaction.\\
\subsection{Positive-$U$ double quantum dot at zero field}
\indent We first consider the conventional case $U>0$ at $B=0$. For $W=0$, the charge configurations $(0,0)$, $(1,0)$, $(0,1)$, and $(1,1)$ meet at a quadruple point satisfying 
\[ E_{00}=E_{10}=E_{01}=E_{11} \quad\Rightarrow\quad (\epsilon_1,\epsilon_2)=(0,0). \]
When $W>0$, the quadruple point splits into two triple points. The first satisfies 
\[ E_{00}=E_{10}=E_{01} \quad\Rightarrow\quad (\epsilon_1,\epsilon_2)=(0,0), \]
while the second satisfies 
\[ E_{10}=E_{01}=E_{11} \quad\Rightarrow\quad (\epsilon_1,\epsilon_2)=(-W,-W). \] 
The separation between these two triple points is \[ \Delta_{01-10}=W. \]
\indent Similarly, the quadruple point involving the charge states $(1,0)$, $(2,0)$, $(1,1)$, and $(2,1)$ is split by $W$. For $W>0$, the resulting triple points are separated along the shifted-energy axis, $\epsilon_{\mathrm S}$, by \[ \Delta_{20-11}=W. \] For $W<0$, the topology of the stability diagram changes and the splitting appears instead along the detuning axis, $\delta$.\\
\subsection{Effect of tunneling and spin-orbit coupling}
\indent Finite interdot tunneling hybridizes charge states and, as a result, the triple points are rounded into avoided crossings. For a two-level system coupled by matrix element $t$, the level repulsion at resonance is $2t$. Spin-orbit coupling introduces an additional spin-flipping tunneling amplitude $t_{\mathrm{SO}}$, leading to an effective tunneling \[ t_{\mathrm{eff}}=\sqrt{t^2+t_{\mathrm{SO}}^2}. \] The corresponding avoided-crossing energies become \[ \Delta_{01-10}=W+2t_{\mathrm{eff}}, \] and \[ \Delta_{20-11}=W+2\sqrt{2}\,t_{\mathrm{eff}}. \] The factor $\sqrt{2}$ arises from the coupling between a paired state and two degenerate spin configurations of the $(1,1)$ manifold.
\subsection{Negative-$U$ double quantum dot}
For $U<0$, the paired charge states $(2,0)$ and $(0,2)$ become energetically favorable. The relevant crossing is therefore between the paired configurations rather than between singly occupied states. In the absence of tunneling, the paired degeneracies satisfy 
\[ E_{00}=E_{20}=E_{02} \quad\Rightarrow\quad (\epsilon_1,\epsilon_2)=\left(-\frac{U}{2},-\frac{U}{2}\right), \] 
and 
\[ E_{20}=E_{02}=E_{22} \quad\Rightarrow\quad (\epsilon_1,\epsilon_2)=\left(-\frac{U}{2}-2W,-\frac{U}{2}-2.W\right). \]
For finite interdot interaction $W$, the corresponding splitting becomes \[ \Delta_{02-20}=2W. \]
Including tunneling hybridizes the $(2,0)$ and $(0,2)$ states and produces the avoided crossing \[ \Delta_{02-20} = 2W-\Lambda + \sqrt{\Lambda^2+4t_{\mathrm{eff}}^2}, \] where \[ \Lambda=\frac{|U|+W}{2}. \]
\\ For magnetic fields exceeding the pairing field, $B>B_{\mathrm P}$, the odd-occupancy states become ground states and the splittings $\Delta_{01-10}$ and $\Delta_{02-11}$ reappear in the charge-stability diagram, similarly to the $U>0$ case. However, at sufficiently large field, the spin states are polarized. Consequently, spin-conserving tunneling $t$ couples the $(0,1)$ and $(1,0)$ configurations, whereas the $(0,2)\leftrightarrow(1,1)$ transition requires spin-flip tunneling and is controlled by $t_{\mathrm{SO}}$. The corresponding splittings are \[ \Delta_{01-10}^{B>B_{\mathrm P}} = W+2t, \] and \[ \Delta_{02-11}^{B>B_{\mathrm P}} = W+2\sqrt{2}\,t_{\mathrm{SO}}. \] These relations form the basis of the extraction of $W$, $t$, and $t_{\mathrm{SO}}$ presented in the main text.

\section{Limits on $\Delta_{02-11}$}
As discussed in the main text and shown in Fig.~3f, no clear peak splitting is observed at the $(0,2)$--$(1,1)$ degeneracy point. One possible explanation is that an anticrossing is present, but that the associated splitting is smaller than the experimental resolution and therefore cannot be directly resolved. Here we consider the maximum splitting compatible with the measured linewidths.\\
\indent We first analyze the resonances away from the degeneracy point, where the peaks are well separated and can be fitted by Lorentzian line shapes,
\begin{equation*} 
L_i(x)=A_i \frac{\gamma_i^2} {(x-x_i)^2+\gamma_i^2}, \qquad i=1,2, 
\end{equation*} 
where $A_i$ is the peak amplitude, $x_i$ is the peak position, and $\gamma_i$ is the half-width at half-maximum (HWHM), corresponding to a full width at half maximum $\mathrm{FWHM}=2\gamma_i$. \\
At the degeneracy point, two nearby resonances contribute to the measured signal. Neglecting possible modifications of the line shape due to hybridization, we approximate the measured response as the sum of two Lorentzians,
\begin{equation*} 
L(x)=L_1(x)+L_2(x), 
\end{equation*} 
If an anticrossing exists, the two resonances are separated by
\begin{equation*} 
\Delta_{02-11}=x_2-x_1. 
\end{equation*} 
Introducing the midpoint 
\begin{equation*} 
x_0=\frac{x_1+x_2}{2}, 
\end{equation*} the individual peak positions can be written as 
\begin{equation*} 
x_1=x_0-\frac{\Delta_{02-11}}{2}, \qquad x_2=x_0+\frac{\Delta_{02-11}}{2}. 
\end{equation*} 
The total line shape becomes
\begin{equation*} 
L(x)= A_1 \frac{\gamma_1^2} {(x-x_0+\Delta_{02-11}/2)^2+\gamma_1^2} + A_2 \frac{\gamma_2^2} {(x-x_0-\Delta_{02-11}/2)^2+\gamma_2^2}. 
\end{equation*} 
For sufficiently small separations, $\Delta_{02-11} \ll \gamma_1,\gamma_2$, the sum is well approximated by a single effective Lorentzian 
\begin{equation*} 
L_{\mathrm{eff}}(x) = A_{\mathrm{eff}} \frac{\Gamma_{\mathrm{eff}}^2} {(x-x_0)^2+\Gamma_{\mathrm{eff}}^2}. 
\end{equation*} 
The height of the effective peak is obtained by evaluating the sum at $x=x_0$, 
\begin{equation*} 
A_{\mathrm{eff}} = A_1 \frac{\gamma_1^2} {\gamma_1^2+\Delta_{02-11}^2/4} + A_2 \frac{\gamma_2^2} {\gamma_2^2+\Delta_{02-11}^2/4}. 
\end{equation*} 
Expanding to second order in $\Delta_{02-11}$ yields 
\begin{equation*} A_{\mathrm{eff}} \simeq (A_1+A_2) - \frac{\Delta_{02-11}^2}{4} \left( \frac{A_1}{\gamma_1^2} + \frac{A_2}{\gamma_2^2} \right). 
\end{equation*} 
Thus the effective peak height decreases quadratically with the peak separation. \\
\indent To determine the effective width, we note that the integrated area of a Lorentzian is
\begin{equation*} 
\int_{-\infty}^{\infty} A_i \frac{\gamma_i^2} {(x-x_i)^2+\gamma_i^2} \, dx = \pi A_i \gamma_i. 
\end{equation*} 
The total area of the two resonances is therefore
\begin{equation*} \mathcal{A} = \pi(A_1\gamma_1+A_2\gamma_2), 
\end{equation*} 
which is independent of $\Delta_{02-11}$. Equating this area to that of the effective Lorentzian gives 
\begin{equation*} 
\Gamma_{\mathrm{eff}} = \frac{A_1\gamma_1+A_2\gamma_2} {A_{\mathrm{eff}}}. 
\end{equation*} 
Substituting the expression for $A_{\mathrm{eff}}$ and expanding to second order in $\Delta_{02-11}$ gives
\begin{equation*} 
\Gamma_{\mathrm{eff}} \simeq \frac{A_1\gamma_1+A_2\gamma_2} {A_1+A_2} \left[ 1+ \frac{\Delta_{02-11}^2} {4(A_1+A_2)} \left( \frac{A_1}{\gamma_1^2} + \frac{A_2}{\gamma_2^2} \right) \right]. 
\end{equation*} 
The effective width therefore increases quadratically with the separation. \\
\indent We note that, in the special case of identical amplitudes and widths, $A_1=A_2=A$ and $\gamma_1=\gamma_2=\gamma$, these expressions simplify to $A_{\mathrm{eff}}\simeq2A\left(1-\Delta_{02-11}^2/4\gamma^2\right)$ and $\Gamma_{\mathrm{eff}}\simeq\gamma\left(1+\Delta_{02-11}^2/4\gamma^2\right)$. In the limit $\Delta_{02-11}\rightarrow0$, one recovers exactly $A_{\mathrm{eff}}=2A$ and $\Gamma_{\mathrm{eff}}=\gamma$, as expected for two perfectly overlapping Lorentzians.\\ 
\indent The amplitudes and linewidths extracted from Lorentzian fits to the resonances away from the degeneracy point (along $\varepsilon$), together with the ones measured at the degeneracy point along the $\varepsilon_\text{s}$ and $\delta$ directions (Fig.~3(f)), are summarized in Table~\ref{tab:lorentzian_parameters}. The table also reports the effective width expected from the overlap of the two resonances assuming $\Delta_{02-11}=0$. We find that this effective width is already comparable to, and even slightly larger, than the linewidth measured at the degeneracy point and even larger than the experimentally observed value. Since any finite splitting $\Delta_{02-11}$ would further broaden the effective resonance, the measured linewidths do not exhibit a statistically significant excess broadening attributable to an unresolved anticrossing. Consequently, within the experimental uncertainty of the present measurements, no meaningful upper bound on the splitting $\Delta_{02-11}$ can be extracted from a linewidth analysis alone. Thus, while a small anticrossing gap cannot be excluded, it remains below the resolution of the experiment.

\begin{table}[H] 
\centering 
\begin{tabular}{|l|c|c|} 
\hline 
Feature & $A$ (2$e^2/h$)& $\gamma$ ($\mu$eV)  \\ 
\hline 
Peak 1 ($\varepsilon$ cut)& 0.49 & 167  \\ 
Peak 2 ($\varepsilon$ cut) & 0.15 & 68  \\ 
Degeneracy point ($\varepsilon_\text{s}$ cut) & 0.67 & 115  \\ 
Degeneracy point ($\delta$ cut) & 0.62 & 98  \\ 
Predicted overlap ($\Delta=0$) & 0.64 & 144  \\ 
\hline 
\end{tabular} 
\caption{Lorentzian fit parameters used in the linewidth analysis. The last row gives the effective width expected for two overlapping peaks in the limit $\Delta=0$.} \label{tab:lorentzian_parameters} 
\end{table}

\section{Comparison of  characteristic energy scales with conventional semiconductor quantum dots}
Table~\ref{tab:hubbard_comparison} summarizes the characteristic energy scales extracted for the present LAO/STO device and compares them with values reported for semiconductor double quantum dots in the literature.

\begin{table}[H] 
\centering 
\begin{tabular}{|l|c|c|c|c|c|c|} 
\hline 
Reference & Material & $U$ (meV) & $W$ (meV) & $t$ ($\mu$eV) & $U/t$ & $W/U$ \\ \hline
This work & LAO/STO & $-0.035$ & 0 & 80 & $-0.4$ & 0 \\ 
\hline 
[38] & GaAs & 0.5 & --- & 10--22 & 10--20 & --- \\ 
\hline 
[39, 40] & Si-MOS & 6.5--11 & 2--3 & 0--400 & 17--$\infty$ & 0.2--0.5 \\ 
\hline
[41] & Si/Ge & 3 & --- & 0-60 & 50--$\infty$ & --- \\ 
\hline 
[42, 40] & Si/SiGe & 4--9 & 2--3 & 90--500 & 8--80 & 0.2--0.8 \\ 
\hline
[43] & Si/SiGe & 6.6 & --- & 5--100 & 66--1320 & --- \\ 
\hline
[44]  & Si/SiGe & 4--8 & 0.9 & 69--517 & 10--80 & 0.2 \\ 
\hline 
[45] & InAs & 5 & 0.5 & 270 & 20 & 0.1 \\ 
\hline 
[46] & Carbon Nanotube & 3--3.5 & --- & 320 & 10 & --- \\ 
 \hline 
\end{tabular} 
\caption{Comparison of energy scales reported for semiconductor double quantum dots and the present LAO/STO device.} 
\label{tab:hubbard_comparison} 
\end{table}

\section{Details of transport simulations}
Transport through the negative-$U$ double quantum dot is modeled using a two-orbital Anderson Hamiltonian implemented in the QmeQ package [36]. The total Hamiltonian consists of the double quantum dot, the leads, and the tunnel coupling between them, \begin{align*} 
H = H_{\mathrm{DQD}} + H_{L} + H_{t}. 
\end{align*} 
The Hamiltonian of the double quantum dot is given by \begin{align*} 
H_{\mathrm{DQD}} &= \sum_{i=1}^{2}\sum_{\sigma} \varepsilon_i n_{i\sigma} + \frac{g\mu_B B}{2} \sum_i \left( n_{i\uparrow}-n_{i\downarrow} \right) \nonumber\\ &\quad + U\sum_i n_{i\uparrow}n_{i\downarrow} + W n_1 n_2 +  \sum_{\sigma\sigma'} d_{1\sigma}^{\dagger} \left[ t\,\delta_{\sigma\sigma'} + i\,t_{\mathrm{SO}}\!\cdot\!\boldsymbol{\sigma}_{\sigma\sigma'} \right] d_{2\sigma'} + \mathrm{h.c.}
\end{align*} where $n_{i\sigma}=d_{i\sigma}^{\dagger}d_{i\sigma}$ is the number operator for electrons with spin $\sigma$ on dot $i$, and $n_i=n_{i\uparrow}+n_{i\downarrow}$ is the total occupation of dot $i$. The single-particle energy of dot $i$ is \begin{align*} \varepsilon_i = \sum_j \alpha_{ij} (V_{\mathrm{G}j} + dE_j), \end{align*} where $\alpha_{ij}$ is the lever arm coupling gate $j$ to dot $i$, $V_{\mathrm{G}j}$ is the applied gate voltage, and $dE_j$ is a voltage offset. The operator $d_{i\sigma}^{\dagger}$ ($d_{i\sigma}$) creates (annihilates) an electron with spin $\sigma$ on dot $i$. The first term in the Hamiltonian describes the single-particle energy of electrons occupying the quantum dots. The second term accounts for the Zeeman splitting induced by an external magnetic field $B$. The third term describes the on-site electron-electron interaction.  The fourth term represents the capacitive (interdot) interaction between electrons occupying different dots. Finally, the last term describes spin-dependent coherent tunneling between the two quantum dots, comprising a spin-conserving hopping amplitude $t$ and a spin-orbit-induced contribution $t_{\mathrm{SO}}$.\\
The Hamiltonian of the leads is given by :
\begin{align*}
H_L=\sum_{\alpha k \sigma} \varepsilon_{\alpha k} c_{\alpha k\sigma}^\dagger c_{\alpha k\sigma} \, ,
\end{align*}
where the operator $c_{\alpha,k}^\dagger$ creates an electron in lead $\alpha$ with momentum $k$ and $\varepsilon_k$ is the corresponding energy. \\
Each lead is coupled to the double quantum dot through the following tunneling Hamiltonian:
\begin{align*}
H_t= \sum_{\alpha k i \sigma} t_{\alpha k i \sigma} c_{\alpha k\sigma}^\dagger d_{i\sigma} +\mathrm{h.c.} \, ,
\end{align*}
where $t_{\alpha,k,i, \sigma}$ is the coupling between state $k, \alpha$ in the lead and state $i, \sigma$ in the quantum dot, which is related to the tunneling rate $\Gamma_\alpha =2\pi \lvert t_{\alpha,k,i, \sigma}\rvert^2$. \\
The conductance through the double quantum dot is calculated using the Pauli master-equation approach implemented in QmeQ, under the assumptions discussed in Ref.~[36]. Within this framework, the leads are assumed to remain in thermal equilibrium and are described by Fermi-Dirac distributions, \begin{align*} f_\alpha(\varepsilon) = \frac{1} {\exp[(\varepsilon-\mu_\alpha)/(k_B T_\alpha)] + 1}, \end{align*} where \(\mu_\alpha\) and \(T_\alpha\) denote the chemical potential and temperature of lead \(\alpha\), respectively. Table~\ref{tab:simulations_parameters} summarizes the parameters used in the simulations. Because the tunnel-coupling parameters $\Gamma_{\alpha}$ were not quantitatively fitted to the experimental data, the absolute conductance scale is arbitrary.

\begin{table}[H]

\begin{center}
    \begin{tabular}{|l|l|}
\hline
Parameter & Value  \\ \hline
$\alpha_{11}$               &    \SI{2.6e-4}{\eV\per\volt}           \\ \hline
$\alpha_{12}$               &    \SI{10e-4}{\eV\per\volt}           \\ \hline
$\alpha_{21}$               &    \SI{0.8e-4}{\eV\per\volt}           \\ \hline
$\alpha_{22}$               &    \SI{1.5e-3}{\eV\per\volt}           \\ \hline
$t$                         &    \SI{8e-5}{\eV}                      \\ \hline
$t_{\mathrm{SO}}$           &    0                      \\ \hline
$U$                         &    \SI{-35e-6}{\eV}                      \\ \hline
$W$                         &    0                      \\ \hline
$g$                         &    \SI{1.5}{}                                \\ \hline
$dE1$                         &    \SI{5e-4}{\V}                         \\ \hline
$dE2$                         &    \SI{1e-3}{\V}                          \\ \hline
$\Gamma_{L,R}$               &     \SI{1.5e-3}{} $t$           \\ \hline
$T_{L,R}$               &      \SI{0.04}{\kelvin}           \\ \hline
\end{tabular}
\end{center}
\caption{Parameters used for the transport simulations presented in the main text.}
\label{tab:simulations_parameters}
\end{table}

\section{Energy diagrams for $U>0$}
\begin{figure}[H]
\centering
\includegraphics[width = \linewidth]{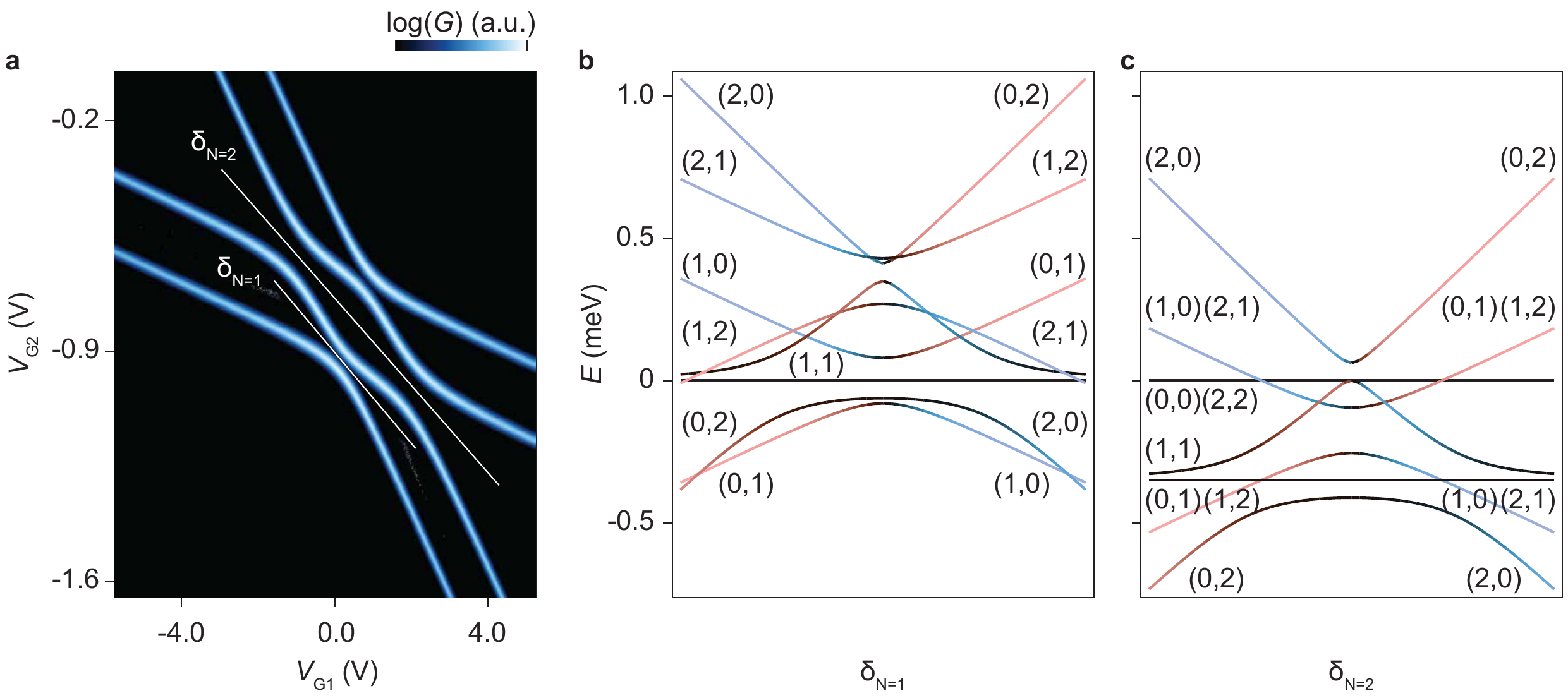}
\caption{ 
\textbf{a} Simulated differential conductance for $U>0$ at $B=0$ T. The dashed lines indicate the two detuning axis $\delta_{N=1}$ and $\delta_{N=2}$. 
\textbf{b} Energy levels along the detuning axis $\delta_{N=1}$. 
\textbf{c} Energy levels along the detuning axis $\epsilon_{N=2}$.
}
\label{S_positiveU}
\end{figure}

Figure \ref{S_positiveU} presents transport simulations and the corresponding energy‑level diagrams along selected detuning axes for the case $U>0$ at $B=0$. Two detuning directions are chosen—one within the total‑charge sector $N_1+N_2=1$  and one within $N_1+N_2=2$, to allow direct comparison with the negative‑$U$ case discussed in the main text.  The parameters used in the simulations are the same as for the negative-$U$ simulations (Table~\ref{tab:simulations_parameters}), except $U=350$~\textmu eV.

